\documentclass[aps,twocolumn,amsmath,amssymb,showpacs,notitlepage]{revtex4-1}
\usepackage[dvipdfmx]{graphicx}
\usepackage{bm}
\usepackage{amssymb}
\usepackage{amsmath}
\usepackage{color}

\begin{document}

\title{Weak coupling theory of topological Hall effect}
\author{Kazuki Nakazawa}
\altaffiliation[Present address: ]{Department of Earth and Space Science, Graduate School of Science, Osaka University, Toyonaka, Osaka 560-0043, Japan}
\altaffiliation[E-mail: ]{nakazawa@spin.ess.sci.osaka-u.ac.jp}
\author{Hiroshi Kohno}
\altaffiliation[E-mail: ]{hkohno@nagoya-u.jp, kohno@st.phys.nagoya-u.ac.jp}
\affiliation{Department of Physics, Nagoya University, Nagoya 464-8602, Japan} 
\begin{abstract}
 Topological Hall effect (THE) caused by a noncoplanar spin texture characterized by a scalar spin chirality 
is often described by the Berry phase, or the associated effective magnetic field. 
 This picture is appropriate when the coupling, $M$, of conduction electrons to the spin texture 
is strong (strong-coupling regime) and the adiabatic condition is satisfied. 
 However, in the weak-coupling regime, where the coupling $M$ is smaller than the electrons' 
scattering rate, the adiabatic condition is not satisfied and the Berry phase picture does not hold. 
 In such a regime, the relation of the effective magnetic field to the spin texture can be 
\lq\lq nonlocal'', in contrast to the \lq\lq local'' relation in the strong coupling case. 
 Focusing on the case of continuous but general spin texture, 
we investigate the THE in various characteristic regions in the weak-coupling regime, 
namely, (1) diffusive and local, (2) diffusive and nonlocal, and (3) ballistic. 
 In the presence of spin relaxation, there arise two more regions 
in the \lq\lq weakest-coupling'' regime: (1$'$) diffusive and local, and (2$'$) diffusive and nonlocal.  
 We derived the analytic expression of the topological Hall conductivity (THC) for each region, 
and found that the condition for the locality of the effective field is governed 
by transverse spin diffusion of electrons. 
 In region 1, where the spin relaxation is negligible and the effective field is local, 
the THC is found to be proportional to $M$, instead of $M^3$ of the weakest-coupling regime. 
 In the diffusive, nonlocal regions (2 and 2$'$), the effective field is given by a spin chirality formed 
by \lq\lq effective spins'' that the electrons see during their diffusive motion. 
 Applying the results to a skyrmion lattice, we found the THC decreses  
as the skyrmion density is increased in region 2$'$, reflecting the nonlocality of the effective field, 
and shows a maximum at the boundary to the \lq\lq local'' region.

\end{abstract}
\date{\today }
\maketitle

\section{Introduction}
Hall effect \cite{Hall} is the phenomena in which an electric current is bent in the transverse direction 
to the applied electric field. 
 While (normal) Hall effect is caused by the Lorentz force due to applied magnetic field, 
anomalous Hall effect (AHE) comes from quantum mechanical effects in materials \cite{Nagaosa}. 
 After the pioneering works \cite{KL, Ye}, various mechanisms that cause AHE have been revealed \cite{Tomizawa, Chen}. 

Nowadays, a non-coplanar spin configuration, which has finite spin chirality, is also known to cause AHE. This phenomenon is often called \lq\lq topological Hall effect (THE)''\cite{Bruno} and has been studied intensively in both theoretically \cite{Ye, Tatara, Bruno} and experimentally \cite{Taniguchi, Neubauer, Kanazawa}. 
 This kind of spin configuration is realized in 
 spin glass systems \cite{Kawamura}, magnetic vortex \cite{Shinjo}, skyrmion systems \cite{Rossler, Muhlbauer}, and so on. 
 Especially, magnetic skyrmions attract considerable attention in spintronics recently, 
and THE can offer an electrical means to probe skyrmions in such materials. 

 Usually THE is discussed in terms of an effective magnetic field due to the Berry phase in real space \cite{Berry}. 
 This is appropriate when the exchange coupling  (which we call $s$-$d$ interaction in this paper) 
between the conduction electrons and localized spins, or the magnetization, is strong (strong-coupling regime), and the conduction electrons adjust their spins to the local spin texture. 
 In this case, the effective magnetic field is determined by the magnetization in a local manner, 
\begin{align}
 B_{{\rm s}, z} ({\bm r}) =  {\bm n} \cdot ( \partial_{x} {\bm n} \times \partial_{y} {\bm n} ) , 
\label{eq:Beff}
\end{align}
where ${\bm n}$, $\partial_{x} {\bm n}$ and $\partial_{y} {\bm n}$ are all at the same point ${\bm r}$. 
 On the other hand, it was shown that the THE occurs even in the weak-coupling regime 
\cite{Tatara, Nakazawa, Denisov}, 
which may be interpreted as due to the nonlocal effective magnetic field. 
  These works focus on three spins, out of many spins distributed discretely, 
which form a scalar spin chirality, ${\bm S}_1 \cdot ({\bm S}_2 \times {\bm S}_3) \ne 0$, 
and do not pay attention to the length scale $q^{-1}$ of the spin texture. 
 On the other hand, if we consider the case of smoothly varying spin texture 
(typically, in ferromagnetic materials), $q^{-1}$ becomes a relevant length scale 
and the relations to and among other parameters 
(mean free path $\ell$, scattering time $\tau$, $s$-$d$ coupling constant $M$, etc.) 
are also important in discussing electron transport phenomena. 
  Depending on these parameters, physical picture of THE will be different, and the topological Hall conductivity (THC) will have different analytic expressions, 
including the form of the effective magnetic field (local or nonlocal). 
  Moreover, these factors are important in interpreting the experimental results even at the qualitative level. 
  In fact, recent experiment on Ce-doped CaMnO$_3$ thin films shows that there is certainly 
a material system for which THE cannot be explained by the standard strong-coupling 
(Berry phase) formula, but the weak-coupling formula can explain 
some of the qualitative features well \cite{Bibes,Nakazawa3}.

 In the course of this study, 
we revisit the work by Onoda, Tatara and Nagaosa (OTN) \cite{Onoda},  
the earliest work that investigated the THE in several regimes. 
 They calculated the THC in the ballistic regime, 
in which the electron mean free path is longer than the characteristic length scale of magnetic structure, 
and observed a crossover between the two mechanisms of THE, 
one based on the Berry phase in real space and the other in momentum space 
\cite{momentum_space}. 
 In this paper, however, we focus on the former, and consider continuous magnetic structures  
with length scale much longer than the lattice constant. 
 We found some disagreement with their result about the locality of the effective field.

 In this paper, we investigate the THE due to the \lq\lq real-space'' spin texture 
focussing on the weak-coupling regime ($M\tau < 1$). 
 We consider a ferromagnetic metal having a general but continuous spin texture. 
 We calculate THC by employing two methods 
for the treatment of general spin textures, 
\lq\lq $u$-perturbation'' 
(or the small-amplitude method \cite{Tatara2}) and \lq\lq $M$-perturbation''. 
 We found that the weak-coupling regime is divided into five characteristic regions, 
as shown in Fig.~\ref{fig:regions}, 
and obtained the analytic expression of THC in each region. 
 As an example, we apply the results to a skyrmion lattice (SkL), 
and found a nonmonotonic dependence of THC on the skyrmion density. 
 This is due to the crossover between the local- and nonlocal-effective-field regimes.

 In a separate paper \cite{Nakazawa3}, we studied the same problem 
by another method (spin gauge field) focusing on the diffusive regime. 
 The present paper is intended to provide a detailed analysis by other two methods 
focusing on the weak-coupling regime. 
 These are complementary to each other and will be useful to grasp overall features  
of the THE in a wide parameter space.

 This paper is organized as follows. 
 In Sec.~\ref{sec:Settings}, we introduce the model and some calculational tools, 
as well as the definition of the five characteristic regions. 
 We calculate THC by the $u$-perturbation method in Sec.~\ref{sec:u},  
and by the $M$-perturbation method in Sec.~\ref{sec:M} for general spin textures. 
 Each method is explained at the beginning of each section. 
 In Sec.~\ref{sec:SkL}, the results are applied to a specific texture, the skyrmion lattice. 
 The results are summarized in Sec.~\ref{sec:results}. 
 Comparison to other method (spin gauge field) and relation to previous studies 
are discussed in Sec.~\ref{sec:Discussion}.

\section{Settings}
\label{sec:Settings}

\subsection{Hamiltonian}

 We consider a free electron system coupled to a continuous spin (or magnetization) texture 
by the exchange interaction, and also subjected to random impurity potential.  
 The Hamiltonian is given by $H = H_{K} + H_{\rm imp} + H_{\rm sd}$,  
\begin{align}
 & H_{K} =
 \int d{\bm r} \, c^\dagger ({\bm r}) 
   \left( - \frac{\hbar^2}{2m} \nabla^2 - \epsilon_{\rm F} \right) c ({\bm r}) , 
\\
 & H_{\rm imp} =
 u_{\rm i} \sum_i \int d{\bm r} \, c^\dagger ({\bm r}) \delta ({\bm r}- {\bm R}_i) c ({\bm r}) , 
\\
 & H_{\rm sd} =
 -M \int {\bm n} ({\bm r}) \cdot ( c^{\dagger} ({\bm r}) \, {\bm \sigma} \, c ({\bm r}) ) , 
\label{eq:Hsd}
\end{align}
where 
$c^{\dagger} = ( c^{\dagger}_{\uparrow}, c^{\dagger}_{\downarrow} )$ are electron creation operators. 
 The first term $H_K$ is the kinetic energy,
with $\epsilon_{\rm F}$ being the Fermi energy. 
 The second term $H_{\rm imp}$ is the coupling to impurities; 
we assume a $\delta$-function potential with strength $u_{\rm i}$  
and at position ${\bm R}_{i}$.  
 The third term $H_{\rm sd}$ represents the exchange interaction to the localized spins, ${\bm n }$, 
where the \lq\lq coupling constant'' 
$M \equiv \hbar/2\tau_{\rm ex}$ is the $s$-$d$ exchange coupling constant multiplied 
by the magnitude of localized spin, with $\tau_{\rm ex}$ being the \lq\lq exchange time''. 
${\bm \sigma} = ( \sigma_{x}, \sigma_{y}, \sigma_{z} )$ are the Pauli matrices.  
 We assume ${\bm n} = {\bm n} ({\bm r})$ is static in time and continuous in space, 
 and work with the Fourier components, 
\begin{align}
 {\bm n} ({\bm r}) &= \sum_{\bm q} {\bm n}_{\bm q} \, {\rm e}^{i {\bm q} \cdot {\bm r}} . 
\label{eq:mag_general}
\end{align}

\subsection{Hall conductivity}

 We study the Hall conductivity on the basis of Kubo formula for the conductivity tensor, 
\begin{align}
 \sigma_{ij} ({\bm Q}, \omega) &= 
\frac{K_{ij}^{\rm R}({\bm Q}, \omega) - K_{ij}^{\rm R}({\bm Q}, 0)}{i\omega}, 
\label{eq:Kubo}
\\
 K_{ij}^{\rm R}({\bm Q}, \omega) &= \frac{i}{V} \int_{0}^{\infty} dt \ {\rm e}^{i(\omega+i0)t} \left< \left[ J_{i}({\bm Q},t), J_{j}({\bm 0}, 0) \right] \right>_H  , 
\end{align}
where  
$ \bm{J} ({\bm Q}) = -e\sum_{\bm{k}} {\bm{v}}_{\bm{k}} c_{\bm{k}-{\bm Q}/2}^{\dagger} c_{\bm{k}+{\bm Q}/2}$ is the Fourier compoent of the current-density operator with wave vector ${\bm Q}$, 
$\bm k$ is the wave vector of conduction electron, 
$\omega$ is the frequency of the applied electric field, 
and $\langle \cdots \rangle_H$ represents equilibrium-statistical as well as quenched-impurity 
averaging. 
 While we consider the response to the uniform electric field (hence the operator $J_{j}({\bm 0}, 0)$ in $K_{ij}^{\rm R}$), 
we retain the wave vector ${\bm Q}$ of the induced current density ($J_{i}({\bm Q},t)$) 
that is supplied by the magnetization texture. 
 The experimentally measurable conductivity of a macroscopic specimen is obtained 
by first taking the uniform limit, ${\bm Q} \to {\bm 0}$, and then the DC limit, $\omega \to 0$.  
 For simplicity, we write as $\sigma_{ij} (\omega)$ for $\sigma_{ij} ({\bm Q}={\bm 0}, \omega)$, 
and as $\sigma_{ij}$ for $\sigma_{ij} (\omega = 0)$.

 In this paper, we consider good metals, and focus on the contribution from the Fermi surface, 
\begin{align}
  \sigma_{ij} ({\bm Q}, \omega)
=\frac{e^2}{2\pi V}\sum_{\bm{k}, \bm{k}'} \langle v_{i} G_{\bm{k}+\bm{Q}/2, \bm{k}'}^{\rm R}(\omega) v'_{j}G_{\bm{k}', \bm{k}-\bm{Q}/2}^{\rm A}(0) \rangle_{\rm imp}. 
\label{eq:sigma_RA}
\end{align}
 Here 
$G_{\bm{k}\bm{k}'}^{\rm R(A)} (\varepsilon) 
= \mp i \int \theta (\pm t) \langle {\rm T} \{ c_{\bm k} (t) ,  c_{{\bm k}'}^\dagger \} \rangle 
  \, e^{-i\varepsilon t} dt $ 
is the exact Green function with full account of $H_K$, $H_{\rm imp}$ and $H_{\rm sd}$. 
 We then make a perturbative expansion with respect to $H_{\rm imp}$ and $H_{\rm sd}$, 
and average over the impurity positions. 
 Within the Born approximation, the (impurity averaged) Green's function is given by
\begin{align}
  G_{{\bm k},\sigma}^{\rm R(A)} (\varepsilon) 
= (\varepsilon - \epsilon_{\bm k} + \sigma M \pm i/2\tau_{\sigma})^{-1}, 
\label{eq:green}
\end{align}
where $\epsilon_{\bm k} = {\bm k}^{2}/2m - \epsilon_{\rm F}$, and 
\begin{align}
 \tau_{\sigma} = \frac{\hbar}{2\pi n_{\rm i} u_{\rm i}^2 \nu_{\sigma} } 
\label{eq:tau}
\end{align}
is the (elastic) scattering time, with $n_{\rm i}$ being the impurity concentration  
and $\nu_{\sigma}$ the density of states (per unit volume) at $\epsilon_{\rm F}$. 
 The subscript $\sigma = \uparrow, \downarrow$ indicates the spin dependence. 
 When the spin dependence is unimportant, as often occurs in the present 
weak-coupling theory, we suppress it and just write as $\tau$, $\nu$, $G_{\bm k}$, etc. 
 
 The Hall conductivity, $\sigma_{ij}^{\rm H}$, is defined by the antisymmetric part of the off-diagonal component of $\sigma_{ij}$, 
\begin{align}
 \sigma_{ij}^{\rm H} = \frac{1}{2} (\sigma_{ij} - \sigma_{ji})  . 
\end{align}
 This observation is crucial in the presence of magnetization texture, as in the present case, 
because it lowers the symmetry of the system and generally induces symmetric components in the off-diagonal elements. 
 In the following, we are interested only in the antisymmetric part of $\sigma_{ij}$, 
and the superscript \lq H' will be suppressed.

\begin{figure}
\hspace*{-8mm}
  \includegraphics[width=100mm]{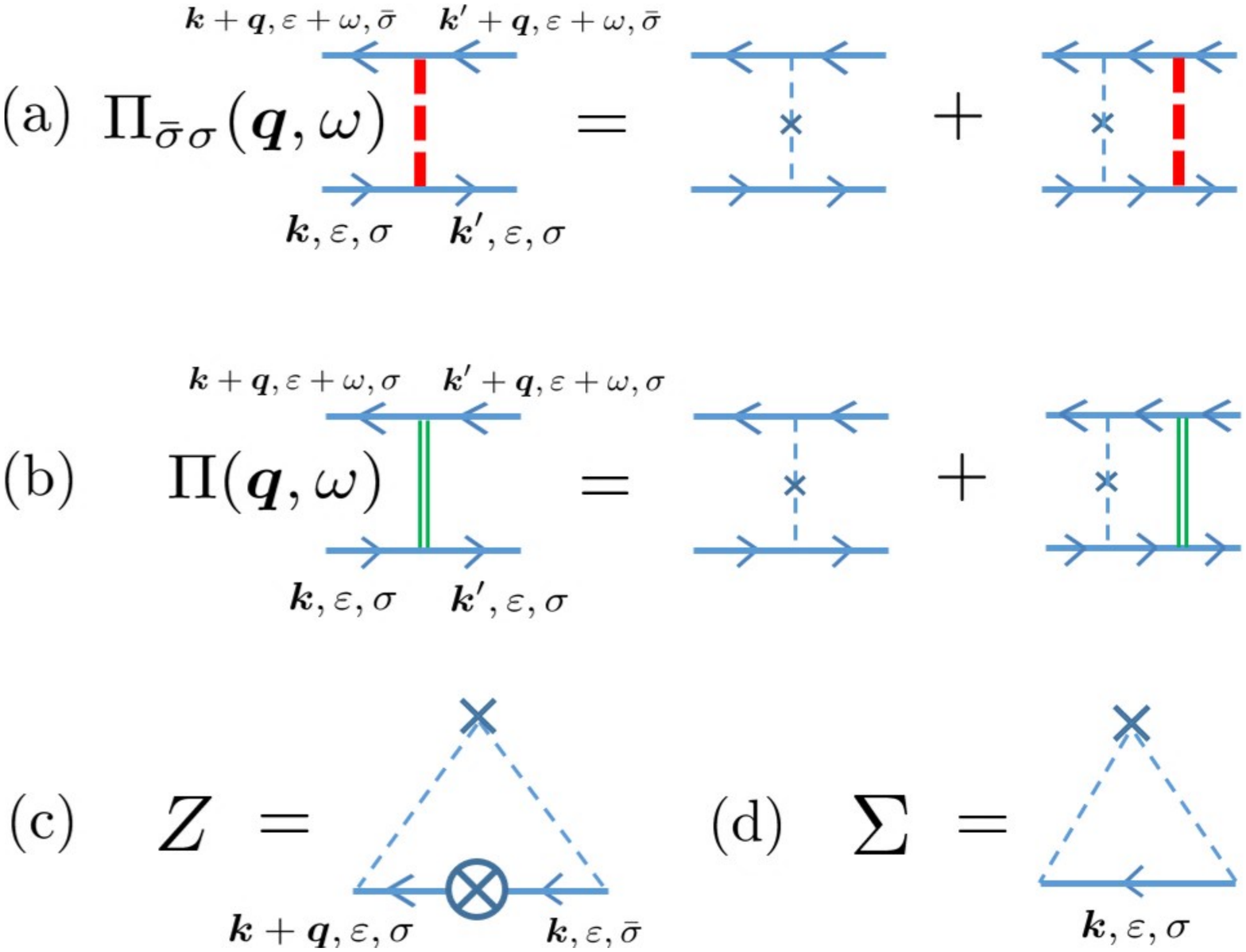}
 \caption{(Color online) 
 Diagrammatic expression of vertex corrections and self-energy due to random impurities. 
 The blue solid line is the Green function of electrons, and 
the thin broken line with a cross represents impurity scattering. 
 (a) Diffusion propagator of transverse spin density, $\Pi_{\bar\sigma\sigma} ({\bm q},\omega)$ 
  (red broken line), which we call $M$-VC. 
 (b) Diffusion propagator of charge density, $\Pi ({\bm q},\omega)$ (green double line), 
 which we call $q$-VC. 
 The upper (lower) lines represent retarded (advanced) Green functions [Eq.~(\ref{eq:green})]. 
(c) Vertex correction to transverse spin, which we call $Z$-VC. 
     The cross with a circle is the (bare) transverse spin vertex. 
(d) Self-energy in the Born approximation.
}
 \label{fig:DPs}
\end{figure}

\subsection{Vertex corrections and diffusion propagator}

 To be consistent with the Born approximation for the self-energy, 
we need to consider the ladder type vertex corrections (VC) for response functions. 

 The scattering between the Green functions with different analyticity (retarded and advanced) 
leads to diffusion propagators (DP). 
 We first introduce DP in the spin channel. 
 It is shown by the red line in Fig.~\ref{fig:DPs}, 
and represents the scattering between the retarded and advanced Green functions with opposite spins. 
 Under the conditions, $q\ell < 1$ and $M\tau < 1$, 
where $\ell \equiv v_{\rm F} \tau$ is the mean free path, it is calculated as   
\begin{align}
  \Pi_{\bar\sigma \sigma} ({\bm q},\omega) 
= \frac{n_{\rm i}u_{\rm i}^2 (1 + 2i\sigma M \tau) }{(Dq^2 + 2i\sigma M - i\omega + \tau_{\rm s}^{-1}) \tau } ,
\label{eq:dp}
\end{align}
with the diffusion constant $D= \frac{1}{3}v_{\rm F}^{2} \tau$ of electrons. 
 We introduced the spin relaxation time $\tau_{\rm s}$ on phenomenological grounds. 
 We call $\Pi_{\bar\sigma \sigma}$ as the $M$-VC or $M$-DP. 
 Physically, it describes the diffusive motion ($\sim Dq^2$) of electrons  
with precession ($\sim M$) and damping ($\sim \tau_{\rm s}^{-1}$) of transverse spin density. 
 We also introduce a DP in charge channel which does not have dependence 
on $M$ and $\tau_{\rm s}$, 
\begin{align}
  \Pi ({\bm q},\omega) = \frac{n_{\rm i}u_{\rm i}^2}{(Dq^2 - i\omega ) \tau} . 
\end{align} 
 We call this $q$-VC or $q$-DP (green double lines in Fig.~\ref{fig:DPs}(b)), 
which physically represents the diffusion of charge density.

  We also consider the scattering between the Green functions with same analyticity 
(retarded and retarded, or advanced and advanced). 
 It is sufficient to consider only the Born type (leading) correction, instead of taking a ladder sum. 
 This is given by
\begin{align}
  Z = n_{\rm i}u^2 \sum_{\bm k} G_{{\bm k},\sigma}^{\rm R} G_{{\bm k},\bar{\sigma}}^{\rm R} 
 \simeq \frac{i}{4\epsilon_{\rm F} \tau} \equiv i\zeta.  
\label{eq:Z-VC}
\end{align}
 We call this as $Z$-VC (Fig.~\ref{fig:DPs} (c)). 
 This $Z$-VC appears at the (transverse) spin vertex, 
and played an important role in the calculation of current-induced spin torque \cite{Kohno}. 
 Below, we will see that this is also the case for THE.

\subsection{Characteristic regions}

\begin{figure}
\hspace*{-11mm}
  \includegraphics[width=149mm]{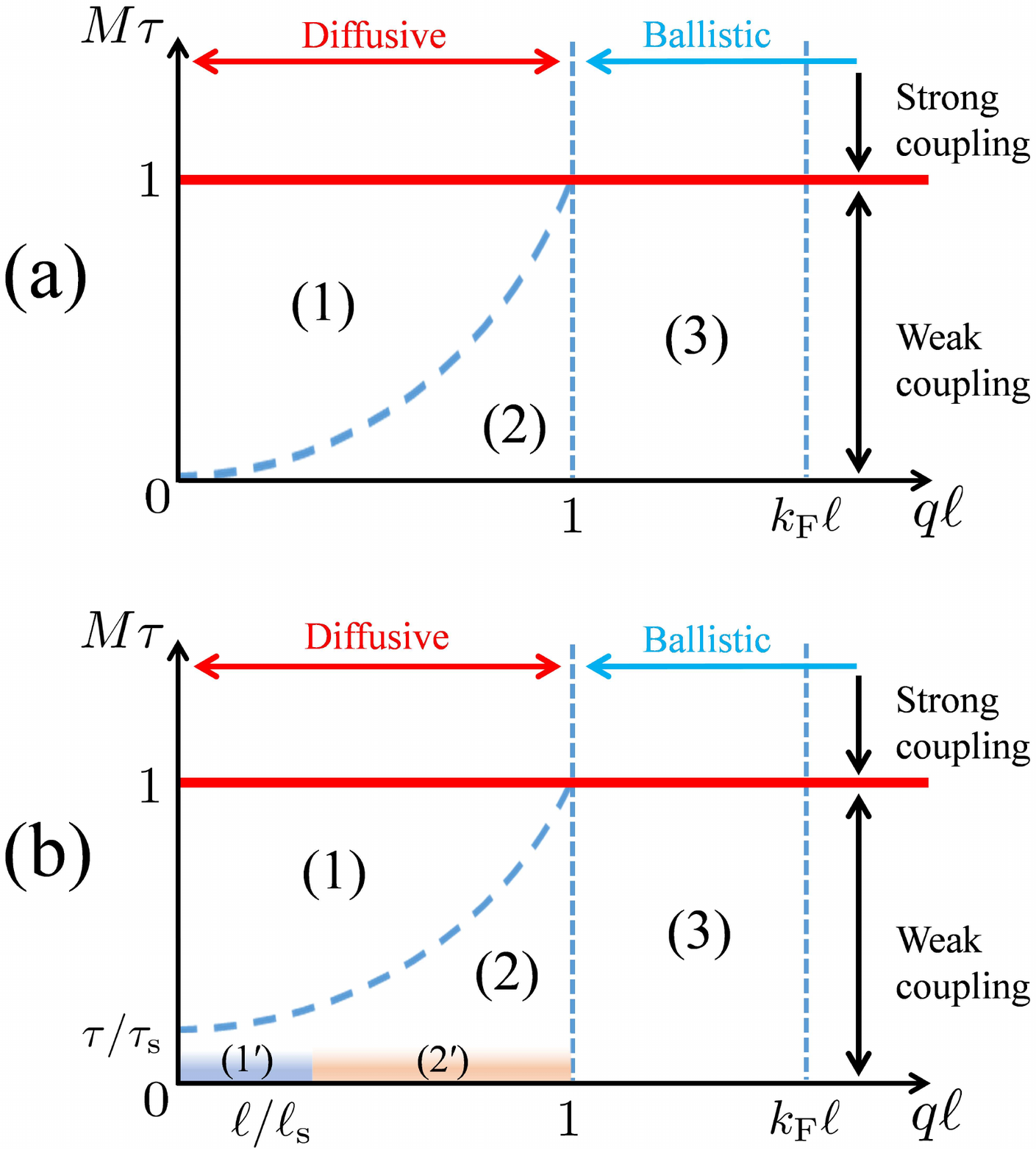}
 \caption{(Color online) Characteristic regions in the weak coupling regime $M\tau <1$ 
 in the plane of $M$ and $q$,   where $2M$ is the exchange splitting, $\tau$ is the electron lifetime, 
 $q$ is the wave vector of spin texture, and $\ell = v_{\rm F} \tau$ is the mean-free path. 
 (a) In the absence of spin relaxation, the weak-coupling regime is divided into three regions; 
 (1) Diffusive and local-effective-field region, 
 (2) Diffusive and nonlocal-effective-field region, and 
 (3) Ballistic region. 
Regions 1 and 2 are separated by the parabola $M\tau = (q\ell)^2$, 
and regions 2 and 3 by the line $q\ell = 1$. 
 (b) In the presence of spin relaxation, the parabola moves to 
$M\tau = (q\ell)^2 + \tau/\tau_{\rm s}$, and there appear two more regions; 
($1'$)  local, and ($2'$) nonlocal. 
}
\label{fig:regions}
\end{figure}

 The qualitative behaviours of the THE are classified into several characteristic regions 
depending on the (relative) values of various parameters. 
 In this paper, we focus on the \lq\lq weak-coupling regime'' defined by 
\begin{align}
  M \tau < 1.
\end{align}
 Also, we restrict ourselves to smooth spin textures, whose characteristic wave vector $q$ satisfies 
\begin{align}
  q \ll  k_{\rm F} , \ a_0^{-1} , 
\end{align}
where $k_{\rm F}$ is the Fermi wave vector and $a_0$ is the lattice constant. 
 This means that the region of momentum-space Berry phase studied in OTN \cite{Onoda} 
is outside the scope of the present paper.

 As a problem of quantum transport, an important length scale whose interplay with $q^{-1}$ is 
essential is the electrons' mean free path $\ell$. 
  This leads to a classification into the \lq\lq diffusive regime'' $(q \ell < 1)$ 
and the \lq\lq ballistic regime'' $(q \ell > 1)$. 
  In the former (diffusive) regime, spatial modulation of the texture is slower than $\ell$,  
and induces only small momentum changes ($q < \ell^{-1}$) of conduction electrons. 
 In the latter (ballistic) regime, the magnetization varies rapidly and involves large momentum transfer ($q > \ell^{-1}$) of electrons.
 In other words,  in the ballistic (diffusive) regime, the electrons see a spin chirality,  
or an effective magnetic field, through their ballistic (diffusive) motion.

 In the diffusive regime, effects of electron diffusion will be important. 
 In fact, we will see that the characteristic behaviour of THC is governed by the transverse spin diffusion propagator, Eq.~(\ref{eq:dp}). 
  In the absence of spin relaxation ($\tau_{\rm s}^{-1}=0$), one sees that 
Eq.~(\ref{eq:dp}) has two characteristic regions, $M \tau > (q \ell)^2$ and $M\tau < (q \ell)^2$. 
In the presence of spin relaxation, two more regions appear,  
$(q \ell_{\rm s})^2 > 1$ and $(q \ell_{\rm s})^2 < 1$.
where $\ell_{\rm s} \equiv \sqrt{ D\tau_{\rm s} }$ is the spin diffusion length. 
 The boundary of these regions is characterized by 
the spin relaxation time $\tau_{\rm s}$
and the so-called Thouless time $\tau_{\rm Th}^{\phantom{\dagger}} = \tau/(q\ell)^2$, 
the time for the electrons to diffuse over the distance of $q^{-1}$.

 To summarize, the THE is expected to show different characteristic behaviours 
depending on the following parameter regions. 
 In the absence of spin relaxation, there are three regions, 
\begin{align}
&{\rm Region \ 1} : \  (q \ell)^2  < M \tau < 1 , \\
&{\rm Region \ 2} : \  M\tau  < (q \ell)^2  < 1  , \\
&{\rm Region \ 3} : \  M\tau < 1 < q\ell . 
\end{align}
  These are shown in Fig.~\ref{fig:regions} (a) in the plane of $M$ and $q$, 
or dimensionless parameters, $M\tau$ and $q\ell$. 
 In the presence of spin relaxation, the following two regions appear 
in the \lq\lq weakest-coupling'' regime $M \tau_{\rm s} < 1$, 
\begin{align}
&{\rm Region \ 1'} : \  M \tau_{\rm s} < 1 , \ \   q \ell_{\rm s} < 1 ,  \\
&{\rm Region \ 2'} : \  M \tau_{\rm s} < 1 , \ \   q \ell_{\rm s} > 1 , 
\end{align}
as shown in Fig.~\ref{fig:regions} (b). 
  In the following sections, we present the calculation and the result of the THC in each region. 
  We use two methods for the treatment of the spatial variation of spin texture; 
the small-amplitude method (\lq\lq $u$-perturbation'')
and a direct perturbative treatment of $H_{sd}$ itself (\lq\lq $M$-perturbation'').

\section{THC studied by $u$-perturbation (Regions 1, 1$'$, 2, and 2$'$)}
\label{sec:u}

\begin{figure}
\hspace*{-18mm}
  \includegraphics[width=122mm]{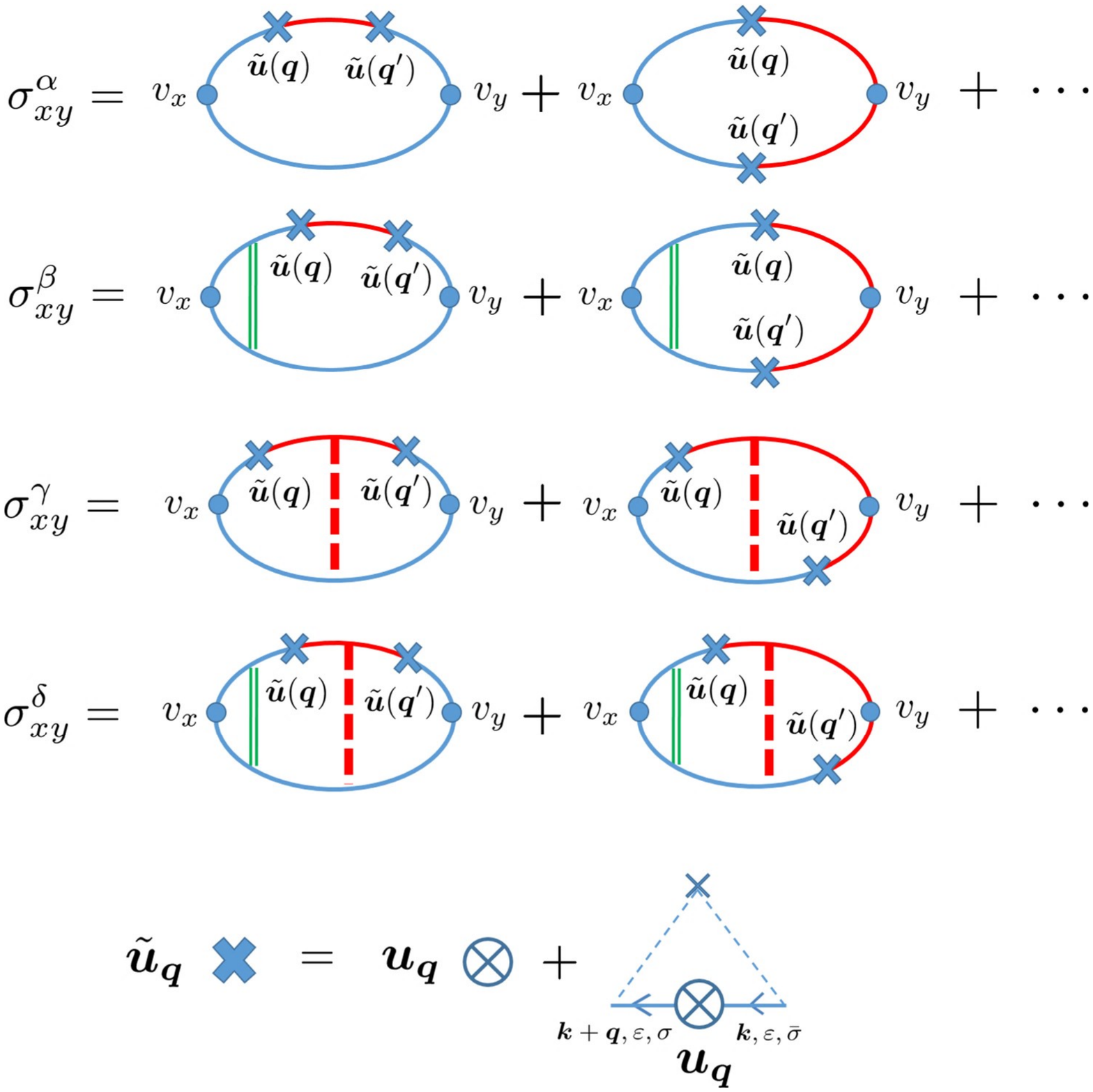}
\vspace{-8mm}
 \caption{(Color online) Feynman diagrams for $\sigma_{xy}$ in the $u$-perturbation method. 
 The blue thick cross ($\tilde {\bm u}_{\bm q}$) represents the \lq\lq $u$-vertex'' which include the $Z$-VC, namely,
 $\tilde {\bm u}_{\bm q} = {\bm u}_{\bm q} ( 1 + i\zeta )$.
 The blue cross with a circle (${\bm u}_{\bm q}$) is the bare $u$-vertex 
given by the second term of Eq.~(\ref{eq:Hsd_u}). 
 The red (blue) solid line represents Green functions of electrons with spin projection $\bar{\sigma}$ ($\sigma$). 
 Of the four groups of the diagrams, in the diffusive regime (regions 1, 1$'$, 2 and 2$'$), 
 $\sigma_{xy}^{\alpha}$ and $\sigma_{xy}^{\beta}$ are irrelevant 
 and both $\sigma_{xy}^{\gamma}$ and $\sigma_{xy}^{\delta}$ are important. 
}
 \label{fig:upert}
\end{figure}

 We first calculate THC using the small amplitude method (\lq\lq $u$-perturbation'')  \cite{Tatara2}. 
 In this method, we consider a small transverse deviation, ${\bm u}$, 
of the magnetization around a uniformly magnetized state, ${\bm n} = \hat{z} $, namely
\begin{align}
 {\bm n} ({\bm r}) = \hat{z} + {\bm u}({\bm r})  , 
\end{align}
with the condition $|{\bm u}| \ll 1$. 
 This method works nicely when the functional form is known beforehand 
and we only need to determine the coefficient. 
 In the present case of THC, we know the functional dependence of $\sigma_{xy}$ on ${\bm n}$, 
which is given by the first equality of 
\begin{align}
  \sigma_{xy} 
&= f ({\bm q},{\bm q}',{\bm q}'') \ {\bm n}_{q''} \cdot (i q_{x} {\bm n}_{q} \times i q'_{y} {\bm n}_{q'} )
\nonumber \\
&\simeq  f ({\bm q},{\bm q}',{\bm 0}) \ (i q_{x} {\bm u}_{q} \times i q'_{y} {\bm u}_{q'} )_{z}.
\label{eq:uu}
\end{align}
 The second equality is obtained as a leading term. 
 Note that ${\bm n}_{\bm q} = \hat z \, \delta_{{\bm q},{\bm 0}} + {\bm u}_{\bm q}$. 
 The coefficients in the first and the second lines are different in their argument, namely, 
$f ({\bm q},{\bm q}',{\bm q}'')$ versus $f ({\bm q},{\bm q}',{\bm 0})$. 
 However, one can compare the results by setting ${\bm q}'' = {\bm 0}$ in the former. 
 Hence, we can obtain THC in the second order with respect to $\bm u$. 
(The whole function $f ({\bm q},{\bm q}',{\bm q}'')$ can be determined by calculating  
 the fourth-order contributions.)
 This is justified even if $M$ is large as far as $u$ is small. 
 We call this method as \lq\lq $u$-perturbation'' in this paper.

 For explicit calculations, we write the $s$-$d$ coupling as 
\begin{align}
H_{\rm sd} 
= \ &
- M \int d^3r \ ( c^{\dagger} ({\bm r}) \sigma^{z} c ({\bm r}) ) \nonumber \\
&- M\int d{\bm r} \ {\bm u} ({\bm r}) \cdot c^{\dagger} ({\bm r}) {\bm \sigma}^{\perp} c ({\bm r}) . 
\label{eq:Hsd_u}
\end{align}
 We include the first term in the unperturbed Hamiltonian 
(hence the term $\sigma M$ in the Green function, Eq.~(\ref{eq:green})) 
and treat the second term perturbatively. 
 Since we are interested in the second-order contribution (see Eq.~(\ref{eq:uu})), 
it is sufficient to consider the magnetic structure of the form, 
\begin{align}
{\bm u} ({\bm r}) = {\bm u}_{ {\bm q} }  {\rm e}^{ i {\bm q} \cdot {\bm r} } + {\bm u}_{ {\bm q}' }  {\rm e}^{ i {\bm q}' \cdot {\bm r} } + {\rm c.c.} 
\end{align}
by retaining two (independent) wave vectors, ${\bm q}$ and ${\bm q}'$. 
 In this section, we set  
\begin{align}
 {\bm Q} &\equiv {\bm q} + {\bm q}' ,  
\end{align}
for the wave vector of the Hall current, see Eqs.~(\ref{eq:Kubo})-(\ref{eq:sigma_RA}).

 In the calculation of THC, first-order perturbation with respect to ${\bm u}$ vanishes, 
consistent with Eq.~(\ref{eq:uu}), and we consider the second-order contribution 
(Fig.~\ref{fig:upert}). 
 We evaluate the diagrams by retaining only the low-order terms with respect to $M$ 
since we are working on the weak-coupling regime $M\tau < 1$. 
 At the same time, we should note that the $M$-VC introduces a \lq\lq singular'' factor, 
having $M$ in the denominator, see Eq.~(\ref{eq:dp}). 
 Generally, the dominant contribution in the diffusive regime ($q\ell <1$) comes from diagrams 
that contain DPs. 
 This is because a DP gives a factor $(q \ell)^{-2}$ relative to those without DP \cite{Nakazawa}. 
 For this reason, we can neglect the diagrams without $M$-VC, 
and we here consider only the diagrams which contain $M$-VC.

 In the following calculation, we first proceed with the general form for the $M$-DP, 
Eq.~(\ref{eq:dp}), which is then specialized to each region. 
 The THE in regions 1, 1$'$, 2, and 2$'$ is essentially given by 
$ \sigma_{xy}^{\rm (1,1',2,2')} \simeq \sigma_{xy}^{\gamma} + \sigma_{xy}^{\delta}$, where 

\begin{widetext}

\begin{align}
 \sigma_{xy}^{\gamma} ({\bm Q}, \omega) 
&= -\frac{e^2}{\pi} M^2 ( {\bm u}_{\bm q} \times {\bm u}_{ {\bm q}' } )^{z}  
\nonumber \\ 
& \quad \times \sum_{\sigma} \sigma \cdot {\rm Im} \left[ X_{ {\bm q}, -{\bm q}' }^{\sigma} (\omega) 
\Pi_{\bar\sigma \sigma} ( { \bm q }', \omega ) \left\{ Y_{{\bm q}'}^{\sigma} (\omega)  - \left( Y_{{\bm q}'}^{\bar{\sigma}} (\omega) \right)^{*} \right\}  
  - ( {\bm q} \leftrightarrow {\bm q}' ) \right]  - (x \leftrightarrow y), 
\\
 \sigma_{xy}^{\delta} ({\bm Q}, \omega) 
&= \frac{e^2}{\pi} M^2 ( {\bm u}_{\bm q} \times {\bm u}_{ {\bm q}' } )^{z} \frac{DQ_{x}}{\tau \left( D Q^2 - i\omega  \right)} 
\nonumber \\
& \quad \times \sum_{\sigma} \sigma \cdot {\rm Re} \left[ V_{ {\bm q}, -{\bm q}' }^{\sigma} (\omega) 
\Pi_{\bar\sigma \sigma}  ( { \bm q }', \omega ) \left\{ Y_{{\bm q}'}^{\sigma} (\omega)  - \left( Y_{{\bm q}'}^{\bar{\sigma}} (\omega) \right)^{*} \right\} 
 - ( {\bm q} \leftrightarrow {\bm q}' )  \right]  - (x \leftrightarrow y) , 
\end{align}
with
\begin{align}
 X_{ {\bm q}, -{\bm q}' }^{\sigma} (\omega) 
&= (1 + i\zeta) \sum_{\bm k} \left( {\bm v} + \frac{ {\bm q} - {\bm q}' }{2m} \right)_{x}  G_{ {\bm k} + {\bm q}, \sigma }^{\rm R} (\omega_{+}) G_{ {\bm k}, \bar{\sigma} }^{\rm R} (\omega_{+}) G_{ {\bm k} - {\bm q}', \sigma }^{\rm A} (\omega_{-}) , 
\\
 V_{ {\bm q}, -{\bm q}' }^{\sigma} (\omega)  
&= (1 + i\zeta) \sum_{\bm k} G_{ {\bm k} + {\bm q}, \sigma }^{\rm R} (\omega_{+}) G_{ {\bm k}, \bar{\sigma} }^{\rm R} (\omega_{+}) G_{ {\bm k} - {\bm q}', \sigma }^{\rm A} (\omega_{-}) , 
\\
 Y_{{\bm q}'}^{\sigma} (\omega)  
&= (1 + i\zeta) \sum_{\bm k} v_{y} G_{ {\bm k} + {\bm q}, \bar{\sigma} }^{\rm R} (\omega_{+}) G_{ {\bm k}, \sigma }^{\rm R} (\omega_{+}) G_{ {\bm k}, \sigma }^{\rm A} (\omega_{-}) ,
\end{align}

\end{widetext}
and ${\bm v} \equiv \hbar {\bm k}/m$ and $\omega_\pm = \pm \omega/2$. 
 Since we are working on the weak-coupling $(M\tau <1)$ 
and diffusive $(q\ell <1)$ regime, 
we can evaluate the integrals, $X$, $Y$ and $V$, by expanding them by the dimensionless parameters 
$M\tau$ and $q\ell$. 
 Deferring the calculation to Appendix A, we give the result as 
\begin{align}
 \sigma_{xy}^\gamma ({\bm Q}, \omega) &=  \sigma_{xy}^\delta ({\bm Q}, \omega) 
\nonumber \\
&= -\frac{4}{9} \left( \frac{e}{m} \right)^2 \nu M^3 \tau^4 \, 
( {\bm u}_{{\bm q}} \times {\bm u}_{ {\bm q}' } )_{z} ( i{\bm q} \times i{\bm q}')^z 
\nonumber \\
& \quad\quad\quad\quad\quad \times 
   \left[  |\Gamma (q)|^2 + |\Gamma (q')|^2 \right]  , 
\label{eq:ugamma} 
\end{align}
where 
\begin{align}
 \Gamma (q) &=  \frac{1}{ ( Dq^2  + \tau_{\rm s}^{-1}  + 2i |M| ) \tau} .  
\end{align}
 The factor $|\Gamma (q)|^2 + |\Gamma (q')|^2$ comes from $M$-VC. 
 This shows that the diagrams with $q$-VC (namely, $\sigma_{xy}^\delta$) gives a comparable 
(actually, the same) contribution with those without $q$-VC (namely, $\sigma_{xy}^\gamma$). 
 Interestingly, this feature is {\it not} shared by the $M$-perturbation 
[see below Eq.~(\ref{eq:chi})].

 The two expansions (with respect to $q\ell$ and $M\tau$) 
done in obtaining Eq.~(\ref{eq:ugamma}) are possible irrespective of the relative magnitude 
of $M\tau$ and $q\ell$ in the Green functions, and are commutative. 
 The behaviour of THC in the weak-coupling ($M\tau <1$) diffusive ($q\ell <1$) regime 
is thus determined by the spin diffusion propagator $\Gamma (q)$. 
 The classification in Fig.~\ref{fig:regions} is based on this observation.
 
 Let us look at the THC in the real-space form, 
\begin{align}
 \sigma_{xy}^{(1,1',2,2')} 
&= - \frac{8}{9} \left( \frac{e}{m} \right)^2 \nu M^3 \tau^4  \, 
\nonumber \\
& \ \   \times 
   \left[ \langle {\bm n} \cdot ( \partial_x {\bm n} \times \partial_y \tilde {\bm n}_0 ) \rangle 
         + \langle {\bm n} \cdot ( \partial_x \tilde {\bm n}_0 \times \partial_y {\bm n} ) \rangle  \right] ,
\label{eq:12_r_u}
\end{align}
where
\begin{align}
  \tilde {\bm n}_0 ({\bm r})  
&=  \frac{1}{8\pi M D \tau^2 } 
   \int d{\bm r}' \ 
   \frac{e^{-a|{\bm r} - {\bm r}'|} \sin (b |{\bm r} - {\bm r}'|) }{ |{\bm r} - {\bm r}'| } \, {\bm n} ({\bm r}') , 
\label{eq:d}
\end{align}
is an \lq\lq effective spin''. 
 The parameters are defined by 
\begin{align}
 a &= \left[ \frac{ \sqrt{ \ell_{\rm s}^{-4} + \lambda^{-4} } + \ell_{\rm s}^{-2}}{2}  \right]^{1/2}, 
\label{eq:a}
\\ 
 b &= (2a\lambda^2)^{-1},  
\label{eq:b}
\end{align}
and 
\begin{align} 
 \lambda = \sqrt{\hbar D/2|M|} = \sqrt{D \tau_{\rm ex}}  , 
\end{align}
is the \lq\lq spin-precession length''.

  An equivalent, but more symmetrical expression can be obtained by writing as 
$|\Gamma (q)|^2 + |\Gamma (q')|^2 = |\Gamma (q) - \Gamma (q')|^2 
 + \Gamma^* (q) \Gamma (q') + \Gamma (q) \Gamma^* (q')$ and neglecting the first term 
(it vanishes in the uniform limit, ${\bm Q} \to {\bm 0}$). 
 Then, 
\begin{align}
 \sigma_{xy}^{(1,1',2,2')}  
&= - \frac{16}{9} \left( \frac{e}{m} \right)^2 \nu M^3 \tau^4  \, 
   {\rm Re} \left[ \langle  {\bm n} \cdot ( \partial_x \tilde {\bm n} \times \partial_y \tilde {\bm n}^* ) \rangle \right] ,
\label{eq:12_r_u}
\end{align}
where 
\begin{align}
& \tilde {\bm n} ({\bm r})  
=   \frac{1}{4\pi D \tau } \int d{\bm r}' \ 
     \frac{e^{-(a+ib) |{\bm r} - {\bm r}'|} }{ |{\bm r} - {\bm r}'| } \,  {\bm n} ({\bm r}') , 
\label{eq:d}
\end{align}
is the effective spin that the electrons see during their diffusive motion. 
 Because of the $q$-dependence of $\Gamma (q)$, $\tilde {\bm n} ({\bm r})$ is related to 
${\bm n} ({\bm r}') $ in a nonlocal way. 
 However, if the spin texture varies slowly compared to $a^{-1}$, this relation becomes a local one.  
 Let us study each case in the following. 

\subsection{ Regions 1 and 2 (weak spin relaxation)}

 When the spin relaxation is weak, $\ell_{\rm s} \gg \lambda$, 
we have $a=b=(\sqrt{2} \lambda )^{-1}$. 
  Namely, the relevant length scale is the spin precession length, $\lambda$, 
and the nonlocality of the relation (\ref{eq:d}) is determined by $q^{-1}$ versus $\lambda$. 
 When $q \lambda < 1$ (region 1), it becomes local, $\tilde {\bm n} = (2iM\tau )^{-1}  {\bm n}$, 
and the THC is given by  
\begin{align}
 \sigma_{xy}^{(1)} 
= -\frac{4}{9} \left( \frac{e}{m} \right)^2 \langle B_{{\rm s}, z} \rangle \, \nu M \tau^2  ,
\label{eq:u_1} 
\end{align}
where $B_{{\rm s}, z}$ is the \lq\lq local'' effective field given by Eq.~(\ref{eq:Beff}), 
and $\langle \cdots \rangle$ represents spatial average. 
 In the opposite case, $q \lambda > 1$ (region 2), it remains nonlocal and the THC is given by 
\begin{align}
 \sigma_{xy}^{(2)} 
&= - \frac{4}{9} \left( \frac{e}{m} \right)^2 \nu M \tau^2  \, 
   {\rm Re} \left[ \langle {\bm n} \cdot (\partial_x \tilde {\bm n}^{(2)} \times \partial_y \tilde {\bm n}^{(2) \, *} ) \rangle \right] ,
\label{eq:u_2}
\\ 
& \tilde {\bm n}^{(2)} ({\bm r})  
= \frac{1}{4\pi \lambda^2}  \int d{\bm r}' \ 
     \frac{e^{- (1+i)|{\bm r} - {\bm r}'|/\sqrt{2}\lambda}}{ |{\bm r} - {\bm r}'|}  \,  {\bm n} ({\bm r}')  . 
\label{eq:2_d}
\end{align}

 The $M$-linear behaviour in region 1 contrasts with the $M^3$-behaviour 
in the \lq\lq weakest-coupling'' regions 1$'$ and 2$'$ (see next). 
 We note that the $Z$-VC plays important roles in giving a correct THC in region 1; 
without it, $\sigma_{xy}^{\gamma}$ vanishes and $\sigma_{xy}^{\delta}$ contains unphysical 
contributions.

\subsection{ Regions 1$'$ and 2$'$ (strong spin relaxation)}

 When the spin relaxation is strong and $\ell_{\rm s}$ becomes comparable to or shorter than 
$\lambda$, namely, $\ell_{\rm s} \underset{\sim}{<} \lambda$, 
we have $a \sim \ell_{\rm s}^{-1}$ and $b \sim \ell_{\rm s}/2 \lambda^2$, 
and the length scale that determines the nonlocality is $\ell_{\rm s}$. 
 When $q \ell_{\rm s} < 1$ (region $1'$), the relation (\ref{eq:d}) becomes local, 
$\tilde {\bm n} = (\tau_{\rm s}/\tau) \, {\bm n}$, and the THC is given by 
\begin{align}
 \sigma_{xy}^{(1')} 
&= - \frac{16}{9} \left( \frac{e}{m} \right)^2 \langle B_{{\rm s}, z} \rangle \, \nu M^3 \tau^2 \tau_{\rm s}^2 . 
\label{eq:1'}
\end{align}
 When $q \ell_{\rm s} > 1$ (region $2'$), the THE is essentially nonlocal,   
\begin{align}
 \sigma_{xy}^{(2')}
&= - \frac{16}{9} \left( \frac{e}{m} \right)^2 \nu M^3 \tau^2 \tau_{\rm s}^2 \, 
     \langle {\bm n} \cdot (\partial_x \tilde {\bm n}^{(2')} \times \partial_y \tilde {\bm n}^{(2')} ) \rangle ,
\label{eq:2'}
\\ 
& \tilde {\bm n}^{(2')} ({\bm r})  
= \frac{1}{4\pi \ell_{\rm s}^2}  \int d{\bm r}' \ 
     \frac{e^{- |{\bm r} - {\bm r}'|/\ell_{\rm s}}}{|{\bm r} - {\bm r}'|} \, {\bm n} ({\bm r}')  . 
\label{eq:2'_d}
\end{align}

\begin{figure}[t]
\hspace*{-18mm}
  \includegraphics[width=122mm]{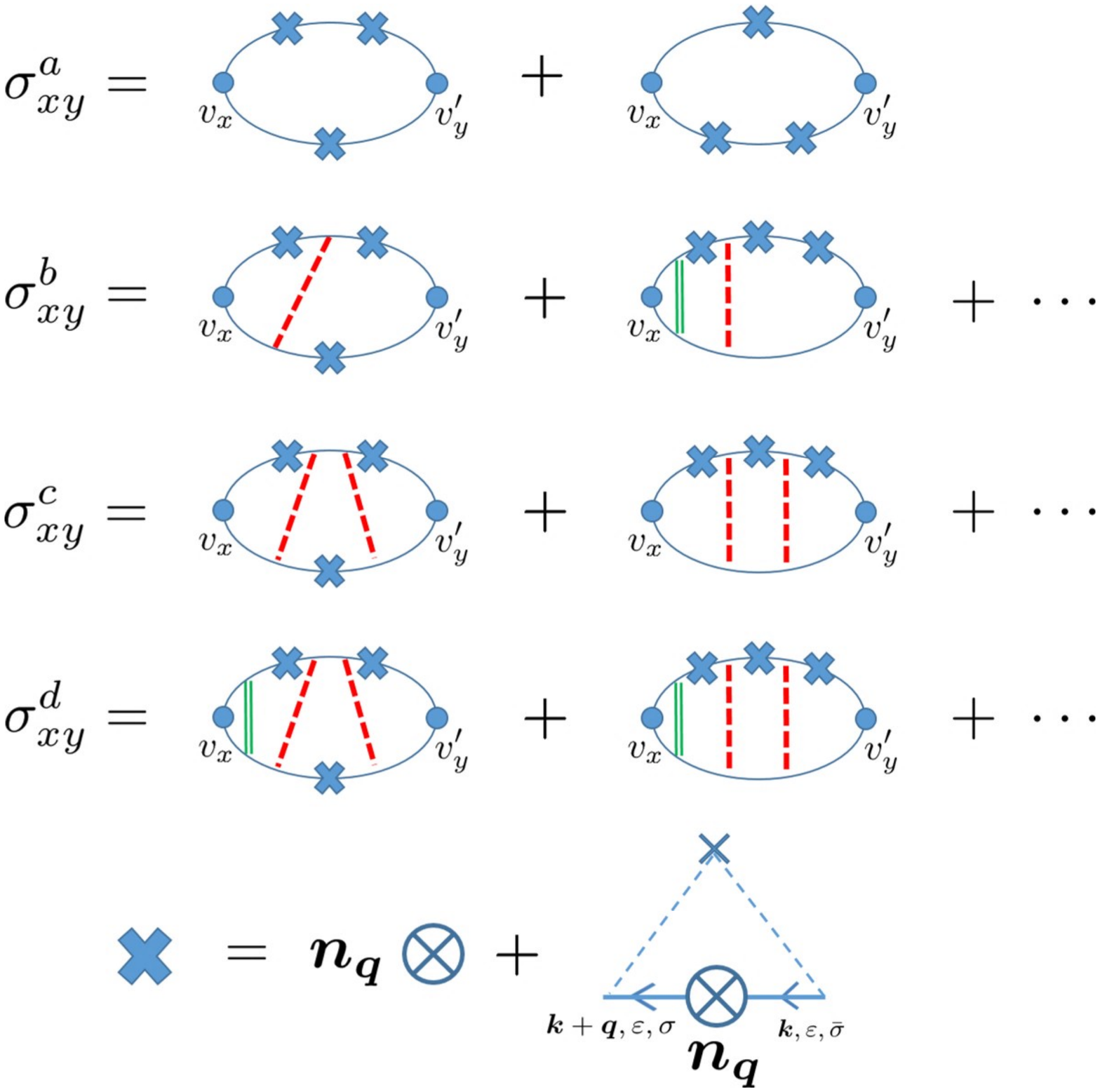}
\vspace{-8mm}
 \caption{(Color online) Feynman diagrams for  $\sigma_{xy}$ in the $M$-perturbation treatment. 
 The thick cross represents the $s$-$d$ coupling, $H_{sd}$, to ${\bm n}$, including the effect of $Z$-VC.  
 The solid lines are electrons' Green functions with $M=0$. 
 The diagrams are classified into 4 groups; $\sigma_{xy}^{a}$ dominantes in region 3, 
and $\sigma_{xy}^{c}$ and $\sigma_{xy}^{d}$ dominante in regions 1$'$ and 2$'$.
}
 \label{fig:Mpert}
\end{figure}

\section{THC studied by $M$-perturbation (Regions 1$'$, 2$'$, and 3)}
\label{sec:M}

 In this section, we calculate THC by treating $M$, or the whole $s$-$d$ coupling $H_{sd}$ [Eq.~(\ref{eq:Hsd})], perturbatively \cite{com1}. 
 This method, which we call \lq\lq $M$-perturbation'', is appropriate for regions 1$'$, 2$'$, and 3, 
where the coupling constant $M$ is the smallest energy scale. 
 Note that this $M$-perturbation may not be justified in regions 1 and 2 even if they belong to 
the weak-coupling region ($M\tau <1$) since $M$ is not the smallest energy scale there. 
 In fact, $M$ is larger than the inverse Thouless time, $\tau_{\rm Th}^{-1} = (q\ell)^2/\tau$ or the inverse spin relaxation time $\tau_{\rm s}^{-1}$.

 For the Hall conductivity, first and second order terms vanish, and the third order term gives a finite contribution. 
  The relevant processes (Feynman diagrams), shown in Figure~4, are similar to those considered 
in the previous studies for a discrete spin distribution \cite{Tatara, Nakazawa}. 
 Here we calculate these diagrams for a continuous spin distribution ${\bm n}$ in each region. 
  Deferring the details to Appendix \ref{sec:M_A}, we describe the outline in this section. 
  It is convenient to retain only three Fourier components,   

\begin{align}
{\bm n} ({\bm r}) &= {\bm n}_{\bm q} {\rm e}^{i {\bm q} \cdot {\bm r}} + {\bm n}_{{\bm q}' } {\rm e}^{i {\bm q}' \cdot {\bm r}} + {\bm n}_{{\bm q}'' } {\rm e}^{i {\bm q}'' \cdot {\bm r}} + {\rm c.c.}  , 
\label{eq:mag}
\end{align}
and focus on the induced Hall current with wave vector, 
\begin{align}
 {\bm Q} &\equiv {\bm q} + {\bm q}' + {\bm q}'' ,
\label{eq:Q}
\end{align}
and the corresponding Hall conductivity, $\sigma_{ij} ({\bm Q}, \omega)$. 
 Considering general ${\bm q}$, ${\bm q}'$ and ${\bm q}''$, 
 it is sufficient to calculate $\sigma_{ij}$ in the first order for each of the three terms in Eq.~(\ref{eq:mag}). 
 One can forget about the complex conjugate part in Eq.~(\ref{eq:mag}) because they do not match 
 the momentum condition (\ref{eq:Q}) in general.

\begin{widetext}
\subsection{Regions 1$'$ and 2$'$}

 The dominant contributions to the THC are given by the diagrams $\sigma_{xy}^{c}$ and $\sigma_{xy}^{d}$ in Fig.~\ref{fig:Mpert}, thus 
$\sigma_{xy} \simeq \sigma_{xy}^{c} + \sigma_{xy}^{d}$.
 The former ($\sigma_{xy}^{c}$) contains two $M$-DPs, 
and the latter ($\sigma_{xy}^{d}$) contains two $M$-DPs and one $q$-DP. 
 They are expressed as 
\begin{align}
 \sigma_{xy}^{c} ({\bm Q}, \omega) 
= &-\frac{4e^2}{\pi} M^3 \chi_{{\bm q},{\bm q}',{\bm q}''} \Pi^{0} ({\bm Q}',\omega) \Pi^{0} ({\bm q}'',\omega)
 {\rm Im} [ \Gamma_{x} ({\bm Q}, {\bm Q}',\omega) ] \cdot
{\rm Im} [ \Gamma ( {\bm Q}', {\bm q}'', \omega) ]  \cdot  {\rm Im} [ \Lambda_{y} ({\bm q}'', \omega) ]  \nonumber \\
&+ ({\bm q} \leftrightarrow {\bm q}') - (x \leftrightarrow y) + ({\rm 2 \ cyclic \ permutations}), 
\\
 \sigma_{xy}^{d} ({\bm Q}, \omega) 
= &-\frac{4ie^2}{\pi} M^3 \chi_{{\bm q},{\bm q}',{\bm q}''}
 \lambda_{x} ({\bm Q},\omega) \Pi({\bm Q},\omega) \Pi^{0} ({\bm Q}',\omega) \Pi^{0} ({\bm q}'',\omega) 
 {\rm Re} [ \Gamma ({\bm Q}, {\bm Q}',\omega) ]  \cdot
{\rm Im} [ \Gamma ( {\bm Q}', {\bm q}'',\omega ) ]  \cdot {\rm Im} [ \Lambda_{y} ({\bm q}'',\omega)  ] \nonumber \\
 &+ ({\bm q} \leftrightarrow {\bm q}') - (x \leftrightarrow y) + ({\rm 2 \ cyclic \ permutations}), 
\end{align}
where ${\bm Q} = {\bm q} + {\bm q}' + {\bm q}''$, ${\bm Q}' = {\bm q}' + {\bm q}''$, and
\begin{align}
  \Gamma ( {\bm q}, {\bm q}',\omega) 
&= (1 + i\zeta) \sum_{\bm k} G_{ {\bm k} + {\bm q} }^{\rm R} (\omega_{+}) 
     G_{ {\bm k} + {\bm q}' }^{\rm R} (\omega_{+}) G_{ {\bm k} }^{\rm A} (\omega_{-})  , 
\\
  \Gamma_{i} ( {\bm q}, {\bm q}',\omega ) 
&= (1 + i\zeta) \sum_{\bm k} \left( {\bm v} + \frac{\bm q}{2m} \right)_{i}  
    G_{ {\bm k} + {\bm q} }^{\rm R} (\omega_{+}) G_{ {\bm k} + {\bm q}' }^{\rm R} (\omega_{+}) G_{ {\bm k} }^{\rm A} (\omega_{-})  , 
\\
  \lambda_{i} ({\bm q},\omega) 
&= \sum_{\bm k} \left( {\bm v} + \frac{\bm q}{2m} \right)_{i}
    G_{ {\bm k} + {\bm q} }^{\rm R} (\omega_{+}) G_{ {\bm k} }^{\rm A} (\omega_{-})  , 
\label{eq:lambda}
\\
  \Lambda_{i} ({\bm q},\omega) 
&= (1 + i\zeta)  \sum_{\bm k} v_{i}  G_{ {\bm k} + {\bm q} }^{\rm R} (\omega_{+}) 
    G_{ {\bm k} }^{\rm R} (\omega_{+}) G_{ {\bm k} }^{\rm A} (\omega_{-}) ,
\end{align}

\end{widetext}
with $\omega_{\pm} \equiv \pm \omega/2$, and 
$\Pi^{0} = \left[ 2\pi \nu \tau^2 (Dq^2 - i\omega + \tau_{\rm s}^{-1} ) \right]^{-1}$ 
is the $M$-VC evaluated at $M=0$.  
 We used 
${\rm tr}[\sigma^{\alpha} \sigma^{\beta} \sigma^{\gamma}] = 2i\varepsilon^{\alpha\beta\gamma}$ 
and defined 
\begin{align}
 \chi_{{\bm q},{\bm q}',{\bm q}''} = {\bm n}_{\bm q} \cdot ( {\bm n}_{{\bm q}'} \times {\bm n}_{{\bm q}''} ) . 
\label{eq:chi}
\end{align}
 Taking first the uniform limit (${\bm Q} \to {\bm 0}$) and then the DC limit ($\omega \to 0$), 
we find $\sigma_{xy}^{d} = 0$. 
Hence, the THC is given by $\sigma_{xy}^{c}$ in both regions 1$'$ and 2$'$, namely, 
$\sigma_{xy}^{(1',2')} \simeq \sigma_{xy}^{c} $. 

 For region 1$'$, we use $\Pi^0 \simeq \tau_{\rm s}/2\pi \nu \tau^2$ to obtain
\begin{align}
 \sigma_{xy}^{(1')} & = - \frac{16}{9} \left( \frac{e}{m} \right)^2 \langle B_{{\rm s}, z} \rangle \, \nu M^3 \tau^2 \tau_{\rm s}^2 . 
\label{eq:m1'}
\end{align}
 For region 2$'$, the $q$-dependence of $\Pi^{0}$ is important, giving  
\begin{align}
  \sigma_{xy}^{(2')} &= -\frac{16}{9} \left( \frac{e}{m} \right)^2 \nu M^3 \tau^2 \tau_{\rm s}^2  \, 
  \langle {\bm n} \cdot ( \partial_{x} \tilde {\bm n}^{(2')} \times  \partial_{y} \tilde {\bm n}^{(2')} ) 
  \rangle , 
\label{eq:m2'}
\end{align}
where
\begin{align}
 \tilde {\bm n}^{(2')} ({\bm r}) & \equiv 
 \frac{1}{V} \sum_{\bm q} \frac{ \ell_{\rm s}^{-2}  }{q^2 + \ell_{\rm s}^{-2} } 
   \, {\bm n}_{\bm q}  \,  {\rm e}^{i {\bm q} \cdot {\bm r}} 
\nonumber \\
&= \frac{1}{4\pi \ell_{\rm s}^2 } 
 \int d{\bm r}' \frac{ e^{- |{\bm r} - {\bm r}'|/\ell_{\rm s} } }{ |{\bm r} - {\bm r}'| } \, {\bm n} ({\bm r}').
\end{align}
 These results agree with those obtained by the $u$-perturbation method [Eqs.~(\ref{eq:1'})-(\ref{eq:2'_d})].

\subsection{Region 3 (Ballistic, nonlocal)}
 
 In region 3, the main contribution to the THC comes from the diagrams without vertex corrections, 
namely, $\sigma_{xy}^{a}$ in Fig.~\ref{fig:Mpert}. 
 For simplicity, we consider the uniform component, $\sigma_{xy}({\bm Q}={\bm 0})$, 
\begin{widetext}
\begin{align}
 \sigma_{xy}^{(3)} &\simeq \sigma_{xy}^{a} 
\nonumber \\  
&= -\frac{ie^2}{\pi V}  M^3
  \sum_{\bm k} \sum_{{\bm q},{\bm q}'} \chi_{{\bm q},{\bm q}',{\bm q}''}
  v_{x} ( {\bm k} + {\bm q}''/2) v_{y} ( {\bm k} - {\bm q}''/2) 
  G_{{\bm k} + {\bm q}''/2}^{\rm R}
  G_{{\bm k} + {\bm q} + {\bm q}''/2}^{\rm R} 
  G_{{\bm k} - {\bm q}''/2}^{\rm R}
  G_{{\bm k} - {\bm q}''/2}^{\rm A} 
  G_{{\bm k} + {\bm q}''/2}^{\rm A} 
\nonumber \\
& \ \ \ +\frac{ie^2}{\pi V}  M^3
  \sum_{\bm k} \sum_{{\bm q},{\bm q}'} \chi_{{\bm q},{\bm q}',{\bm q}''}
  v_{x} ( {\bm k} - {\bm q}''/2) v_{y} ( {\bm k} + {\bm q}''/2) 
  G_{{\bm k} - {\bm q}''/2}^{\rm R} 
  G_{{\bm k} + {\bm q}''/2}^{\rm R} 
  G_{{\bm k} + {\bm q}''/2}^{\rm A} 
  G_{{\bm k} - {\bm q} - {\bm q}''/2}^{\rm A} 
  G_{{\bm k} - {\bm q}''/2}^{\rm A}  , 
\label{eq:zero}
\end{align}
\end{widetext}
where ${\bm q}'' = - {\bm q} - {\bm q}'$. 
 After some calculations, we obtain 
\begin{align}
  \sigma_{xy}^{\rm (3)} 
= \frac{2e^2}{m} \nu M^3 \tau^2  \sum_{{\bm q},{\bm q}',{\bm q}''}  \langle \chi_{{\bm q},{\bm q}',{\bm q}''} 
  \, {\rm e}^{i {\bm Q} \cdot {\bm r}}\, {\rm Re} \Phi_{z} ({\bm q},{\bm q}',{\bm q}'') \rangle .
\label{eq:formula}
\end{align}
 The explicit form of $\bm \Phi$ is given by Eqs.~(\ref{eq:Phi})-(\ref{eq:f-f}) in Appendix B.
 Unfortunately, it is quite complicated, e.g., the function in Eq.~(\ref{eq:F-f}) 
depends on the angle between ${\bm q}$ and ${\bm q}'$. 
  This means that the Hall conductivity depends on the very details of the spin texture. 
 It also depends on the band structure if we go beyond the parabolic dispersion. 
 Even the $\tau$-dependence depends on the details of the texture 
 (and the band structure). 
 In the next section, we apply this formula to a skyrmion lattice (triple-$q$ state), 
and obtain a simple behaviour, $\sigma_{xy}^{(3)} |_{\rm SkL} \propto M^3 \tau^2$.

\section{Case of skyrmion lattice}
\label{sec:SkL}

 So far we considered a continuous but general spin texture. 
 Although analytic expressions are available, that for region 3 is quite complicated, 
and it is difficult to grasp the overall feature across the different regions 
in the whole weak-coupling regime.

 Here, we consider a particular texture called skyrmion lattice focusing on the 
\lq\lq weakest-coupling'' regime (regions 1$'$, 2$'$ and 3), 
that range from the diffusive to ballistic regimes. 
 We consider a \lq\lq triple-$q$ state'' expressed by the superposition of three helices \cite{Muhlbauer}, 
\begin{align}
 {\bm M}(\bm r) 
&= M  \sum_{\ell = a, b, c} \left[ \hat{{\bm z}} \cos{( {\bm q}_{\ell} \cdot {\bm r} )} 
 + ( \hat{\bm q}_{\ell} \times \hat{\bm z} )\sin{( {\bm q}_{\ell} \cdot {\bm r} )} \right] 
\nonumber \\
&= \sum_{\ell = a, b, c} \left[ {\bm M}_{\ell} {\rm e}^{i {\bm q}_{\ell} \cdot {\bm r}} 
    + {\bm M}_{\ell}^{*} {\rm e}^{-i {\bm q}_{\ell} \cdot {\bm r}} \right]  , 
\label{eq:M_sk}
\end{align}
where ${\bm q}_{\ell}$ and $M$ are wave vectors and the (common) amplitude of the helices 
(multiplied by the exchange coupling constant, hence the same one as in the previous sections), 
and ${\bm M}_{\ell} \equiv M \{ \hat{z} + i ( \hat{\bm q}_{\ell} \times \hat{\bm z} ) \}$. 
 The wave vectors satisfy 
\begin{align}
|{\bm q}_{a}| = |{\bm q}_{b}| = |{\bm q}_{c}| \equiv q \label{eq:allQ} ,  \\
{\bm q}_{a} + {\bm q}_{b} + {\bm q}_{c} = {\bm 0}  . 
\end{align}
 Because of these relations, the complicated general expression for region 3 is greatly simplified. 

 In the diffusive regime (regions 1$'$ and 2$'$), we obtain 
[Eq.~(\ref{eq:THC_skL_I})]
\begin{align}
  \sigma_{xy}^{\rm (1',2')} |_{\rm skL} 
= -16 \left( \frac{e}{m} \right)^2 \nu M^3 \tau^4  B_{{\rm skL}, z}^{(1',2')} , 
\label{eq:M_1'2'}
\end{align}
where 
\begin{align}
  {\bm B}^{\rm (1',2')}_{\rm skL} 
&= \frac{ 1 }{\left\{ (q\ell)^2 + (\ell/\ell_{\rm s}) \right\}^2} 
\sum_{\ell, m, n} {\rm Re}[ \chi_{\ell m n} ]
\left( i{\bm q}_{\ell} \times i{\bm q}_{m}  \right) 
\nonumber \\
&\sim \frac{18 q^2 }{ \ell^4 \left( q^2 + \ell_{\rm s}^{-2} \right)^2 } 
   ( \hat{{\bm q}}_{a} \times \hat{{\bm q}}_{b} ) . 
\label{eq:sk_Beff}
\end{align}  
 The THC in this case depends on the wave number $q$ of the helices 
and the spin diffusion length $\ell_{\rm s}$, but not on the scattering time $\tau$.

 In the ballistic regime (region 3), we obtain [Eqs.~(\ref{eq:THC_skL_II}) and (\ref{eq:THC_skL_III})]
\begin{align}
  \sigma_{xy}^{\rm (3)} |_{\rm skL} 
&=  -\frac{8\sqrt{3}\pi}{9} \left( \frac{e}{m} \right)^2 \nu M^3 \tau^4 B_{{\rm skL}, z}^{(3)} , 
\end{align}
where 
\begin{align}
  {\bm B}^{\rm (3)}_{\rm skL} 
&=  \sum_{\ell,m,n} {\rm Re}[ \chi_{\ell,m,n} ]  \frac{ ( i {\bm q}_{\ell} \times i {\bm q}_{m} ) }{(q\ell)^2 } 
= 18 \frac{ ( \hat{{\bm q}}_{a} \times \hat{{\bm q}}_{b} ) }{\ell^2} . 
\end{align}
 Interestingly, the THC is proportional to $M^3 \tau^2$, 
but does not depend on the helical pitch $q^{-1}$ (or the skyrmion size).

\begin{figure}
 \begin{center}
  \includegraphics[width=80mm]{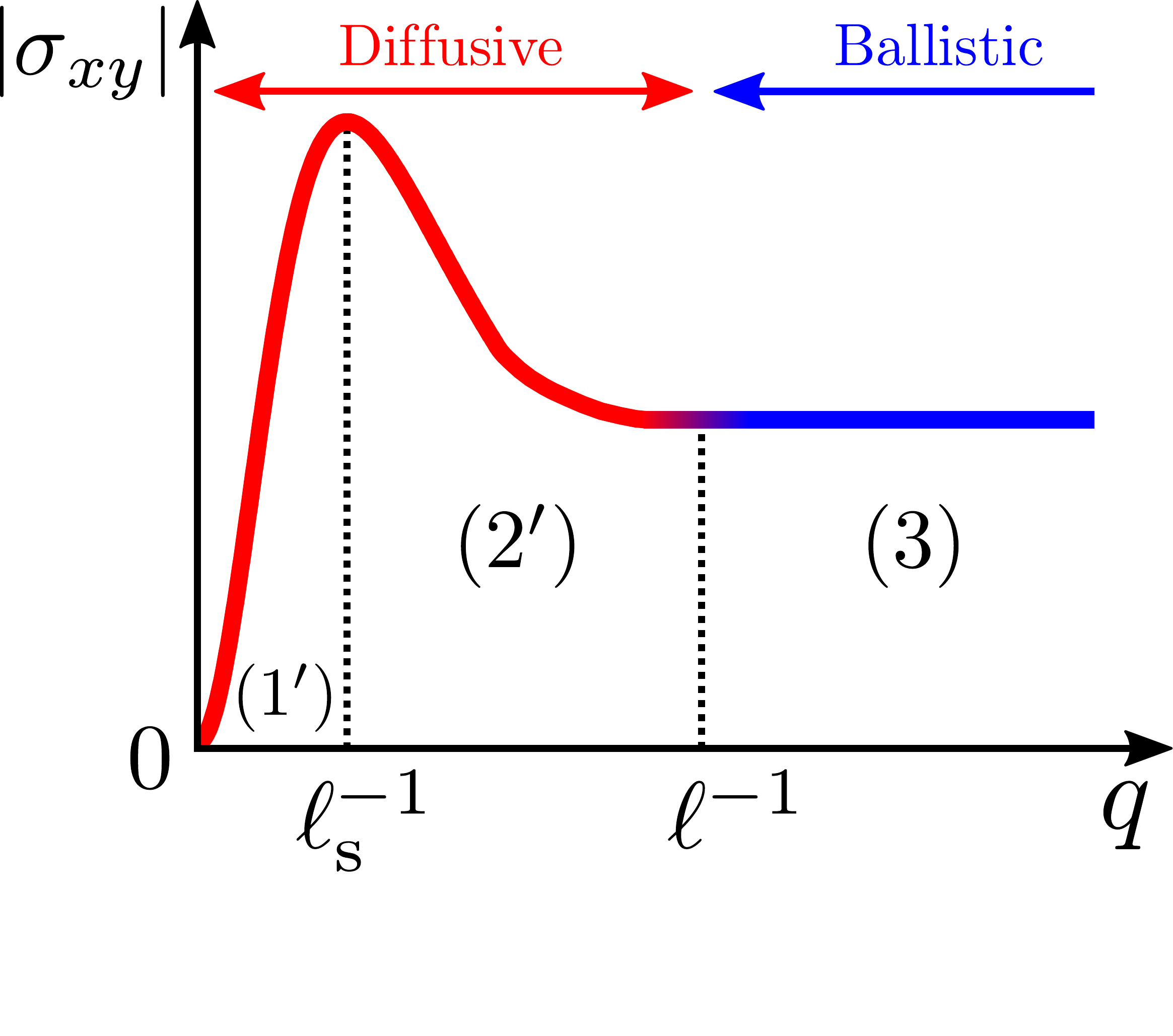}
 \end{center}
\caption{(Color online) Schematic behaviour of THC in a skyrmion lattice 
as a function of the wave number $q$ of the magnetic texture (helix). 
 The THC is larger for lower skyrmion density in region 2$'$, 
a behavior opposite to the strong-coupling regime, 
and takes a maximum at the boundary of regions 1$'$ and 2$'$. 
}
 \label{fig:sk}
\end{figure}

 Figure \ref{fig:sk} shows a schematic plot of the THC as a function of $q$ for fixed $\ell$ 
 (e.g., for a fixed impurity concentration). 
 In region 1$'$, the effective magnetic field is \lq local', 
and the THC monotonically increases as the skyrmion density ($\propto q^2$) is increased. 
 As $q$ is increased further and enters region 2$'$ ($q\ell_{\rm s} > 1$),  
the THC turns to decrease as $\sim q^{-2}$, showing a maximum at around the boundary of 
regions 1$'$ and 2$'$. 
 This behavior in region 2$'$ is opposite to that in the strong-coupling regime as well as in region 1$'$. 
 This is because the effective magnetic field is nonlocal in region 2$'$. 
 If the skyrmions become smaller (and its density higher), electrons see distant spins 
compared to the skyrmion size due to the diffusive motion and the effective spin 
is reduced because of the increased degree of cancellation. 
 In the ballistic regime (region 3), the THC is independent of $q$.

 A similar behaviour is obtained for a scan through regions 1, 2 and 3, 
which appear in the weakest-coupling regime when the spin relaxation is negligible. 
 To see this, we use Eq.~(\ref{eq:u_2}) obtained by the $u$-perturbation method 
since the $M$-VC, which is essential in regions 1 and 2, requires to go 
beyond the low-order $M$-perturbation. 
 In this case, the crossover length scale is given by the spin precession length $\lambda$ 
(instead of $\ell_{\rm s}$). 
 The THC formula is given by Eqs.~(\ref{eq:M_1'2'}) and (\ref{eq:sk_Beff}), 
with the denominator $(q^2 + \ell_{\rm s}^{-2})^2$ in Eq.~(\ref{eq:sk_Beff}) 
being replaced by $(q^2 + \ell_{\rm s}^{-2})^2 + \lambda^{-4} \sim q^4 + \lambda^{-4}$.

 Such local/nonlocal crossover may give rise to some interesting phenomena. 
 As an illustration, consider a system that contains equal number of skyrmions and anti-skyrmions 
but their sizes are different, e.g., with large skyrmions and small anti-skyrmions. 
 (Spin textures with coexisting skyrmions and anti-skyrmions are known to be stabilized in some cases 
\cite{Okubo,Lin}.) 
 In the local regime, the THC is determined by the total topological charge, 
hence it vanishes for this texture because of the cancellation of the contributions from 
skyrmions and anti-skyrmions. 
 However, if the system enters the nonlocal regime, the THC can be nonzero. 
 This happens  when the crossover length scale 
(shorter of $\ell_{\rm s}$ and $\lambda$) 
lies between the two length scales of the texture (one for skyrmions and one for anti-skyrmions). 
 In this case, the effective field is nonlocal for the smaller-size texture (e.g., anti-skyrmions), 
thus reduced from the value of the local case, 
whereas it is local for the larger-size texture (e.g., skyrmions).  
 Thus the cancellation of the two opposite-sign contributions from skyrmions and anti-skyrmions 
is incomplete, and a finite value of the THC will result.

 In this context, we note that there is (at least) one more example that a Hall effect 
is caused by a (dynamical) spin texture with no net spin chirality.  
 According to Ref.~\cite{Yamamoto}, a Hall effect can arise from the scattering of electrons 
by dipolar magnons, whose spatial correlation has no net spin chirality, 
but only a pseudo chirality.

\section{Summary of the results}
\label{sec:results}

\subsection{Results}

 The results obtained in this paper are summarized as follows. 

\noindent
$\bullet$ Region 1 \ (Diffusive, Local)
\begin{align}
 \sigma_{xy}^{(1)} &= -\frac{4}{9} \left( \frac{e}{m} \right)^2 \langle B_{{\rm s}, z} \rangle \, \nu M \tau^2 . 
\end{align}

\noindent
$\bullet$ Region 2 \ (Diffusive, Nonlocal) 
\begin{align}
& \sigma_{xy}^{(2)} 
= - \frac{4}{9} \left( \frac{e}{m} \right)^2 \nu M \tau^2  \, 
   {\rm Re} \left[ \langle {\bm n} 
   \cdot ( \partial_x \tilde {\bm n}^{(2)} \times \partial_y \tilde {\bm n}^{(2) \, *}) \rangle \right] ,
\\ 
& \ \ \ \  \tilde {\bm n}^{(2)} ({\bm r})  
= \frac{1}{4\pi \lambda^2}  \int d{\bm r}' \ 
     \frac{e^{- (1+i)|{\bm r} - {\bm r}'|/\sqrt{2}\lambda}}{ |{\bm r} - {\bm r}'|}  \,  {\bm n} ({\bm r}')  . 
\label{eq:n_2}
\end{align}

\noindent
$\bullet$ Region 1$'$ \ (Diffusive, Local, Strong spin relaxation)
\begin{align}
 \sigma_{xy}^{(1')} 
&= -\frac{16}{9} \left( \frac{e}{m} \right)^2 \langle B_{{\rm s}, z} \rangle \, \nu M^3 \tau^2 \tau_{\rm s}^2 . 
\end{align}

\noindent
$\bullet$ Region 2$'$ \ (Diffusive, Nonlocal, Strong spin relaxation)
\begin{align}
& \sigma_{xy}^{(2')} 
= - \frac{16}{9} \left( \frac{e}{m} \right)^2 \nu M^3 \tau^2 \tau_{\rm s}^2  \, 
  \langle {\bm n} \cdot ( \partial_{x} \tilde {\bm n}^{(2')} \times \partial_{y} \tilde {\bm n}^{(2')} ) \rangle , 
\\ 
& \ \ \ \  \tilde {\bm n}^{(2')} ({\bm r}) = 
\frac{1}{4\pi \ell_{\rm s}^2 } 
 \int d{\bm r}' \frac{ e^{- |{\bm r} - {\bm r}'|/\ell_{\rm s} } }{ |{\bm r} - {\bm r}'| } \, {\bm n} ({\bm r}'). 
\label{eq:n_2'}
\end{align}

\noindent
$\bullet$ Region 3 \ (Ballistic, Nonlocal)
\begin{align}
 \sigma_{xy}^{(3)} &= \frac{2e^2}{m} \nu M^3 \tau^2 \, \sum_{{\bm q}, {\bm q}', {\bm q}''} 
  \langle  \chi_{{\bm q},{\bm q}',{\bm q}''} \, {\rm e}^{i {\bm Q} \cdot {\bm r}} 
  \, {\rm Re} \, \Phi_{z}  \rangle . 
\end{align}

 In regions 1 and 1$'$, the effective magnetic field is local, and the THC depends on $M$ 
like $\sim M$ (region 1) and $\sim M^3$ (region 1$'$). 
 In regions 2 and 2$'$, the effective field is nonlocal, but the expression of THC  
can be simplified if expressed by the effective spin that the electrons see during their diffusive motion. 
 The local/nonlocal boundary is determined by the wavelength $q^{-1}$ of magnetic texture 
relative to the spin precession length $\lambda$ or the spin diffusion length $\ell_{\rm s}$, 
whichever is smaller.

  In region 3, the effective field is nonlocal, and the expression of THC is quite complicated. 
 In this (ballistic) region, the diffusion propagator is not important and a simple bubble diagram 
is sufficient, 
but the details of the band structure and Fermi surface shape can be important. 
 On the other hand, the THC in the diffusive regime does not depend on the details of the band structure.

Table \ref{tab:table} summarizes the characteristic features of THC in each region. 
 Note that the $\tau$ dependence in region 3 is for a specific case of the skyrmion lattice; 
for general textures, it depends on the details of the texture.

\begin{table}[t]
\newlength{\height} 
\setlength{\height}{3.6mm}
	\begin{tabular}{|c||c|c|c|c|} \hline
\rule{0cm}{\height}	Region  &  \ B/D  \  &  Locality & \, $M$-dependence \, & \ $\tau$-dependence \ \\ \hline \hline
\rule{0cm}{\height}	1 & D	&	Local		   &	$M^1$		&	$\tau^2$		\\ \hline	 
\rule{0cm}{\height}	2 & D	& Nonlocal	& \multicolumn{2}{|c|}{ $M^2 \tau F (\lambda q)$}  \\ \hline
\rule{0cm}{\height}	1$'$ &	D	&	Local	  &	$M^3$		&	$\tau^2 \tau_{\rm s}^2$ \\ \hline	 	
\rule{0cm}{\height}	2$'$		&	D	&	Nonlocal	&	$M^3$ &	 
 $\tau \tau_{\rm s} \tilde F (\ell_{\rm s} q)$	\\ \hline
\rule{0cm}{\height}	3 & B	 &	\ Nonlocal \	& $M^3$  & 	$\tau^2$ (SkL) \\ \hline 
	\end{tabular}
\caption{
 Characteristic features of THC in each region, that include 
 \lq\lq ballistic'' (B) or \lq\lq diffusive'' (D), locality of the effective field, 
and the dependence on $M$, $\tau$ and $\tau_{\rm s}$. 
$F(\lambda q)$ and $\tilde F(\ell_{\rm s} q)$ are scaling functions 
with $q^{-1}$ being a single characteristic length scale of the spin texture.
 The $\tau$-dependence in region 3 is for a special case of skyrmion lattice (SkL). 
}
\label{tab:table}
\end{table}

\begin{figure}[t]
\vspace{0cm}
\hspace*{-8mm}
  \includegraphics[width=98mm]{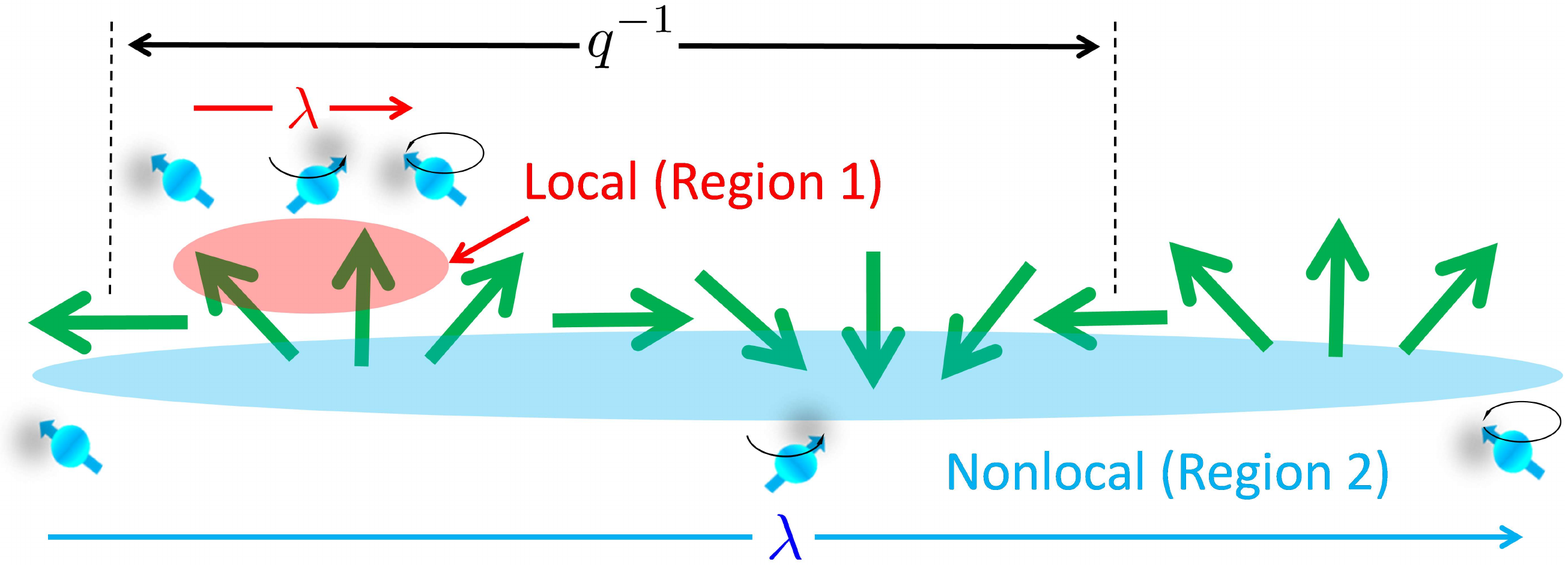}
\vspace{-38mm}
 \caption{ (Color online)  Physical picture of region 1 where the effective field is local (upper), 
and region 2 where the effective field is nonlocal (lower). 
  The thick green arrows represent the magnetization, 
 and the (blue) tiny arrow with a sphere represents the spin of the conduction electron. 
  During the single period of precession $\tau_{\rm ex} = \hbar/2|M|$, electrons diffuse over the distance  
$\lambda = \sqrt{D\tau_{\rm ex}}$ shorter (longer) than the characteristic texture size $q^{-1}$ 
for region 1 (for region 2), and see a small portion of (a wider region than) the texture. 
}
\label{fig:pic}
\end{figure}

\subsection{On the locality of effective field}

 The results for regions 1 and 1$'$ indicate that, even in the weak-coupling regime, 
the Hall conductivity is proportional to the total topological charge 
(or precisely, its density $\langle B_{{\rm s}, z} \rangle$) 
if the effective field is local. 
 This fact allows us to call the Hall effect studied here as \lq\lq topological'' 
even in the weak-coupling regime. 
 On the other hand, in the nonlocal regime, such topological character is diminished. 
(However, we would like to continue to use the words \lq\lq topological'', THE, THC, etc., 
also in this case.)
 Being free from the topological constraint, the THC in the nonlocal regime can be finite 
even when the overall topological charge is zero. 
 A simple example has been given at the end of Sec.~V.

 A physical picture behind the (non-)locality of the effective field 
is illustrated in Fig.~\ref{fig:pic}. 
 As stated, it is governed by the competition between the diffusion ($\sim Dq^2$) 
and the precession ($\sim M$), or the relaxation ($\sim \tau_{\rm s}^{-1}$), of the electron spin, 
described by the diffusion propagator, 
\begin{align}
  \Pi_{\bar{\sigma}\sigma} ({\bm q}, \omega=0) \propto \frac{1}{Dq^2 +2i\sigma M + \tau_{\rm s}^{-1}} ,   
\end{align}
of the transverse spin density [see Eq.~(\ref{eq:dp})].

 Consider first the case that the spin relaxation is weak ($M \tau_{\rm s} > 1$) 
and focus on regions 1 and 2. 
 In region 1 ($Dq^2 \ll M$ or $\lambda \ll q^{-1}$), the spin precession is \lq\lq fast'' in the sense that  
during a single period of precession the electrons diffuse only over a small portion of the texture. 
 This leads to the local expression of the effective magnetic field. 
 On the other hand, in region 2 ($Dq^2 \gg M$ or $\lambda \gg q^{-1}$), the spin precession 
is \lq\lq slow'' 
and the electrons diffuse farther beyond the typical size of the texture in the spin precession 
period. 
 This leads to the nonlocal effective field.

 When the spin relaxation is strong ($M \tau_{\rm s} < 1$) and the spin diffusion length $\ell_{\rm s}$ 
is shorter than $\lambda$, the relevant length scale that determines the nonlocality is $\ell_{\rm s}$. 
 The explanation in the preceding paragraph still holds 
if $\lambda$ is replaced by $\ell_{\rm s}$, 
the \lq\lq spin precession'' by \lq\lq spin relaxation'', 
and the \lq\lq spin precession period'' by \lq\lq spin relaxation time''. 
 This case applies to regions 1$'$ and 2$'$.

\subsection{Analysis of nonlocal effective field }

 When the effective field is nonlocal, the THC does not in general show simple power law dependence 
on $M$ and/or $\tau$. 
 Here we give a scaling-like analysis that works well when there is only one characteristic length scale 
in the spin texture.

 Let us start with region 2. 
 By a change of variables, ${\bm r} = \lambda {\bm s}$, where ${\bm s}$ is dimensionless, 
Eq.~(\ref{eq:n_2}) becomes 
\begin{align}
 \tilde {\bm n}^{(2)} (\lambda {\bm s})  
= \frac{1}{4\pi}  \int d{\bm s}' \ 
     \frac{e^{- (1+i)|{\bm s} - {\bm s}'|/\sqrt{2}}}{ |{\bm s} - {\bm s}'|}  \,  {\bm n} (\lambda {\bm s}')  . 
\end{align}
 We assume that the spin texture ${\bm n}({\bm r})$ has only one length scale $q^{-1}$. 
 Then $\tilde {\bm n}^{(2)} (\lambda {\bm s})$ depends on $q$ and $\lambda$ through 
$\lambda q$, and by dimensional analysis, the effective field should have the form, 
\begin{align}
  {\rm Re} \left[ \langle {\bm n} 
   \cdot ( \partial_x \tilde {\bm n}^{(2)} \times \partial_y \tilde {\bm n}^{(2) \, *}) \rangle \right] 
& = \lambda^{-2} F(\lambda q) , 
\label{eq:Beff_2}
\end{align}
where $F$ is a \lq scaling function'. 
 This relation (and the function $F$) also applies to the local-effective-field region (region 1), 
$\lambda q < 1$, where 
$F(\lambda q) \sim \lambda^2 \, \langle B_{{\rm s}, z} \rangle \sim (\lambda q)^2$.  
 Note that the local effective field $\langle B_{{\rm s}, z} \rangle$ scales as $\sim q^2$. 
 For $\lambda q \gg 1$ (but under the diffusive condition, $q\ell <1$), it decays as 
$F(\lambda q) \sim (\lambda q)^{-2}$. 
 The THC is expressed as 
\begin{align}
& \sigma_{xy}^{(1,2)} 
= - \frac{8}{9} \left( \frac{e}{m} \right)^2 \nu M^2 \tau  \, \frac{ F(\lambda q) }{(D/\tau)} , 
\end{align}
which applies to regions 1 and 2. 
 The scaling function $F$ have a peak as a function of $q$. 
 This is consistent with the example of skyrmion lattice presented in Sec.~V, 
and gives a generalization of it.

 The analysis for region 2$'$ (together with region 1$'$) can be done similarly, 
with $\lambda$ being replaced by $\ell_{\rm s}$. 
 The THC is given by 
\begin{align}
  \sigma_{xy}^{(1',2')} 
&= - \frac{16}{9} \left( \frac{e}{m} \right)^2 \nu M^3 \tau \tau_{\rm s} \, 
       \frac{\tilde F(\ell_{\rm s} q)}{(D/\tau)} . 
\end{align}
where $\tilde F (\ell_{\rm s} q)$ is another scaling function (but similarly 
behaves as $\tilde F(x) \sim x^2$ for $x < 1$, 
and $\tilde F(x) \sim x^{-2}$ for $1 \ll x < \ell_{\rm s}/\ell$).   
 These results are also shown in Table I.

 We emphasize that the analysis in this subsection is based on the assumption that the spin texture 
is characterized by a single length scale. 
 For textures having several characteristic length scales $\{ q_1^{-1}, q_2^{-1}, \cdots \}$, 
one needs to use scaling functions with several variables, 
$F (\lambda q_1, \lambda q_2, \cdots)$ and $\tilde F (\ell_{\rm s} q_1, \ell_{\rm s} q_2, \cdots)$.

\section{Discussion}
\label{sec:Discussion}

In this section, we discuss the results obtained in this paper by comparing with those 
based on the spin gauge field and try to figure out the THE in the whole parameter range. 
 Discussion is also given in relation to the previous studies in the literature.

\subsection{Comparison with gauge field method \cite{Nakazawa3}}

 In Ref.~\cite{Nakazawa3}, we used the spin gauge field to derive the THC formula 
for the weak-coupling, diffusive regime (regions 1, 1$'$, 2 and 2$'$) as 
\begin{align}
 \sigma_{xy}^{\perp, M} 
&= - \frac{16}{9} \left( \frac{e}{m} \right)^2 \nu M^3 \tau^4  \, 
   {\rm Re} \left[ \langle {\bm n} \cdot ( \tilde {\bm d}_{x} \times \tilde {\bm d}_{y}^* ) \rangle \right] ,
\label{eq:12_r_A}
\\ 
 \tilde {\bm d}_{i} ({\bm r})  
&=   \frac{1}{4\pi D \tau } 
     \int d{\bm r}' \ 
     \frac{e^{-(a+ib) |{\bm r} - {\bm r}'|} }{ |{\bm r} - {\bm r}'| } \nonumber \\ 
&\quad \quad \quad \quad \quad \quad \quad \times {\cal R}({\bm r}) \, {\cal R}^{-1} ({\bm r}') \, \partial_{i} {\bm n} ({\bm r}') , 
\label{eq:dt}
\end{align}
where $a$ and $b$ are given by Eqs.~(\ref{eq:a}) and (\ref{eq:b}), and 
${\cal R} ({\bm r}) $ is an SO(3) matrix that satisfies 
${\cal R} ({\bm r}) \hat{z} = {\bm n}({\bm r})$. 
 These are to be compared with Eqs.~(\ref{eq:12_r_u}) and (\ref{eq:d}). 
 There is a difference that comes from the presence of the SO(3) matrix 
${\cal R}({\bm r}) \, {\cal R}^{-1} ({\bm r}')$ in Eq.~(\ref{eq:dt}). 
 In fact, if the factor ${\cal R}({\bm r}) \, {\cal R}^{-1} ({\bm r}')$ is absent, we can write as 
$\tilde {\bm d}_{i} = \partial_i \tilde {\bm n}$ and the two results coincide.  
 The difference disappears in the \lq local' regions, 1 and 1$'$, 
since ${\cal R}({\bm r}) \, {\cal R}^{-1} ({\bm r}') = 1$ at ${\bm r}={\bm r}'$, 
but persists in the \lq nonlocal' regions, 2 and 2$'$.

 The gauge-field result is actually consistent with the present $u$-perturbation result 
since the extra factor, 
${\cal R}({\bm r}) \, {\cal R}^{-1} ({\bm r}')$, or its deviation from unity, is higher order in ${\bm u}$. 
 (One needs to look at terms of ${\cal O}(u^4)$ to fix this.)  
 This is, however, not the case when comparison is made with the $M$-perturbation result. 
 The disagreement is serious in region 2$'$, where the $M$-perturbation is reliable. 
 On the other hand, strictly speaking, the gauge field method is valid in the small $q$ limit, 
but in regions 2 and 2$'$, there are longer length scales, 
$\lambda$ and $\ell_{\rm s}$, than $q^{-1}$, which might invalidate the gauge field result  
(at the simplest level). 
 Therefore, the result by the $M$-perturbation will be the correct one in region 2$'$. 
 As for region 2, no methods are strictly valid at present. 

 Except for the above point, the results obtained by the three methods are consistent, 
thus revealing the overall features in the weak-coupling regime.

\subsection{Comparison with OTN \cite{Onoda}}

 OTN studied both the real-space Berry phase regime and momentum-space Berry phase regime. 
 Here, we focus on the former regime, which corresponds to our region 3. 
 In this region, they obtained the local expression for the effective magnetic field, 
whereas our result indicates the nonlocal effective field. 
 Let us discuss about this discrepancy.

 The argument by OTN for this case is essentially based on the formula obtained 
by Tatara and Kawamura (TK) \cite{Tatara} for the ballistic and weak-coupling regime, 
\begin{align}
 \sigma_{xy} 
&= \sum_{ {\bm r}_1, {\bm r}_2 ,  {\bm r}_3 } 
  K( {\bm r}_1, {\bm r}_2, {\bm r}_3) \, 
  {\bm S} ({\bm r}_{1}) \cdot \left( {\bm S} ({\bm r}_{2}) \times {\bm S} ({\bm r}_{3}) \right) , 
\label{eq:TK}
\end{align} 
where ${\bm r}_{1}$, ${\bm r}_{2}$ and ${\bm r}_{3} $ are the positions of localized spins. 
 The kernel $K( {\bm r}_1, {\bm r}_2, {\bm r}_3)$ contains the RKKY-type factor, 
$I(r_{ij}) = (\sin{k_{\rm F}r_{ij}}/k_{\rm F}r_{ij}) \, {\rm e}^{-r_{ij}/2\ell }$, 
and its derivative, $I' = dI/dr_{ij}$, 
and decays as a function of the separations, $r_{ij} =|{\bm r}_i - {\bm r}_j|$, 
with the length scale given by the mean free path $\ell$. 
 OTN considered the case of continuous magnetic texture and took a continuum limit, 
$\sum_{ {\bm r}_i } \to \int d{\bm r}_i$ and 
\begin{align}
&{\bm S} ({\bm r}_{1}) \cdot \left( {\bm S} ({\bm r}_{2}) \times {\bm S} ({\bm r}_{3}) \right) 
\nonumber \\
&\simeq  ({\bm a} \times {\bm b})_z \, {\bm S} ({\bm r}_{2}) \cdot 
             \left( \partial_{x} {\bm S} ({\bm r}_{2}) \times \partial_{y} {\bm S} ({\bm r}_{2}) \right)  , 
\label{eq:cont}
\end{align}
where ${\bm a}= {\bm r}_{1} - {\bm r}_{2}, \ {\bm b}= {\bm r}_{2} - {\bm r}_{3}$. 
 They obtained 
\begin{align}
\sigma_{xy}^{\rm OTN} \propto M^3 \tau^2 A \int d{\bm r} \ {\bm n} \cdot ( \partial_{x} {\bm n} \times  \partial_{y} {\bm n} )  , 
\label{eq:s_xy_OTN}
\end{align}
with the \lq\lq coefficient''
\begin{align}
 A \sim \int^{q^{-1}} da \int^{q^{-1}}  db \ a^2 b^2 I'(a) I'(b) I(|{\bm a}+{\bm b}|) . 
\label{eq:A_OTN}
\end{align}
 This expansion is justified when the spin texture varies slowly compared to the range 
of the kernel $K$, i.e., $q^{-1} > \ell$, namely, for the diffusive regime, $q \ell <1$ \cite{TYO}.  
 However, in the diffusive regime, this contribution, Eq.~(\ref{eq:s_xy_OTN}), without VC 
is less dominant compared to those with VC as we have seen in Sec.~IV.

 On the other hand, OTN \lq derived' the above formula for $q \ell >1$, but obviously 
the expansion is not justified since the spin texture varies rapidly within the range of the kernel $K$. 
 In fact, the coefficient $A$ in Eq.~(\ref{eq:A_OTN}) depends on $q^{-1}$ and is not simply a constant. 
  It is in fact a complicated function of $q$, hence should enter in the integrand 
of Eq.~(\ref{eq:s_xy_OTN}) as a nonlocal kernel. 
 This means that we need to accept the nonlocal expression (\ref{eq:TK}) as it is. 
 This is consistent with our observation that the effective field is nonlocal in region 3.

\subsection{Comparison with Denisov et al. \cite{Denisov}}

 Denisov, Rozhansky, Averkiev and L\"ahderanta (DRAL) \cite{Denisov} 
considered the case that (small) skyrmions are 
distributed dilutely in a 2D sheet, and calculated THC in two steps. 
 First, they solve the scattering problem on a single skyrmion in the second Born approximation  
and showed that it induces a skew scattering as if there is an effective magnetic field $\Omega$ 
that couples to the electron charge. 
 Because of the diluteness of skyrmions, 
  the problem is similar to the ordinary low-field Hall effect, and they obtained 
\begin{align}
 \sigma_{xy}^{\rm DRAL} = &\ \sigma_{0} \frac{\Omega \tau}{1 + (\Omega \tau)^2} , 
\ \ \ \  \Omega \propto n_{\rm sk} M^3 , 
\label{eq:denisov} 
\end{align} 
for THC, where $\sigma_{0}$ is the Drude conductivity, $n_{\rm sk}$ is the skyrmion sheet density, 
and $M$ is the coupling constant between conduction electrons and localized spins. 
 The leading contribution is proportional to $M^3 \tau^2$ in the weak-coupling regime, 
which agrees with the perturbative calculation by TK \cite{Tatara}. 
 This is because they assume a large momentum change at the scattering 
from the magnetic moments and this is in common with TK. 
 In Appendix D, we give a brief calculation that verifies this.

 The spin texture considered by DRAL is characterized by two length scales, 
one is the size of a skyrmion and the other is the distance, $r_{\rm sk}$, between skyrmions. 
 The former corresponds to our $q^{-1}$ and the latter is a new independent parameter. 
 DRAL considered the case $q^{-1} < \ell < r_{\rm sk}$. 
 On the other hand, we assumed smooth spin textures that satisfy $q^{-1} = r_{\rm sk}  < \ell$ for region 3, 
and $\ell  < q^{-1} = r_{\rm sk} $ for other regions. 
  From the viewpoint of momentum transfer $q$, their situation corresponds to our region 3 
(ballistic regime).

\vspace{0.7cm}

\section{Conclusion}
\label{sec:conclusion}

 We have investigated the topological Hall effect 
in the weak-coupling regime ($M\tau < 1$). 
 Assuming a general but continuous magnetic structure, we calculated the THC in five characteristic regions 
(1, 1$'$, 2, 2$'$, and 3) classified by electron transport properties (diffusive, ballistic), 
locality of the effective magnetic field (local, nonlocal), and the degree of spin relaxation. 
 We obtained the analytic expression of THC for each region. 
  In regions 1 and 1$'$, the effective field is determined by the local magnetic structure, 
and the THC shows an $M$-linear and $M$-cubic dependence, respectively. 
 In regions 2 and 2$'$, the effective fields are nonlocal but have simple forms if expressed 
by \lq\lq effective spins''. 
 In region 3, the effective field is nonlocal and the THC has a complicated expression.

 We then applied the results to a skyrmion lattice (triple-$q$ state), and found that the THC takes a maximum 
at the boundary between regions 1$'$ and 2$'$ (or 1 and 2), namely, at the local/nonlocal boundary 
of the effective magnetic field.  
 In region 2$'$ (and 2), because of the nonlocality, 
the THC increases as the size of the skyrmions is increased (density is decreased),  
a behavior opposite to the \lq\lq local'' regime.

 Even in the weak-coupling regime, the THC is determined 
by the total topological charge (total spin chirality) if the effective field is local, 
whereas such topological constraint is relaxed when the effective field becomes nonlocal. 
 A scaling argument has been given to analyze the THC in the nonlocal region.

 Experimentally, the characteristic $M$-linear behaviour found in region 1 
may have relevance to Ca$_{1-x}$Ce$_x$MnO$_3$ thin films \cite{Bibes}.
 In fact, an estimate indicates that this material is located in region 1 \cite{Bibes}. 
 
\section*{acknowledgement}

 We gratefully acknowledge Manuel Bibes for sharing with us his insights into our theoretical 
result in relation to his and his collaborators' experimental results.
 We also thank J. Fujimoto, T. Yamaguchi and J. Nakane for valuable discussion. 
 This work is supported by JSPS KAKENHI Grant Numbers 25400339, 15H05702 and 17H02929. 
KN is supported by Grant-in-Aid for JSPS Research Fellow Grant number 16J05516, and by a Program for Leading Graduate Schools ``Integrative Graduate Education and Research in Green Natural Sciences''.

\begin{widetext}

\appendix

\section{Calculation based on $u$-perturbation}
\label{sec:u_a}

 In this Appendix, we present some details of the calculation of THC based on the $u$-perturbation 
method. 
 Relevant diagrams are shown in Fig.~\ref{fig:upert}, which are  
$\sigma_{xy}^{\alpha}$, $\sigma_{xy}^{\beta}$, $\sigma_{xy}^{\gamma}$ and $\sigma_{xy}^{\delta}$. 
 We first calculate the dominant contribution ($\sigma_{xy}^{\gamma}$ and $\sigma_{xy}^{\delta}$), 
and then show that $\sigma_{xy}^{\alpha}$ and $\sigma_{xy}^{\beta}$ are less dominant.  

\begin{figure}
\vspace{0cm}
 \begin{center}
 \includegraphics[width=175mm,pagebox=cropbox,clip]{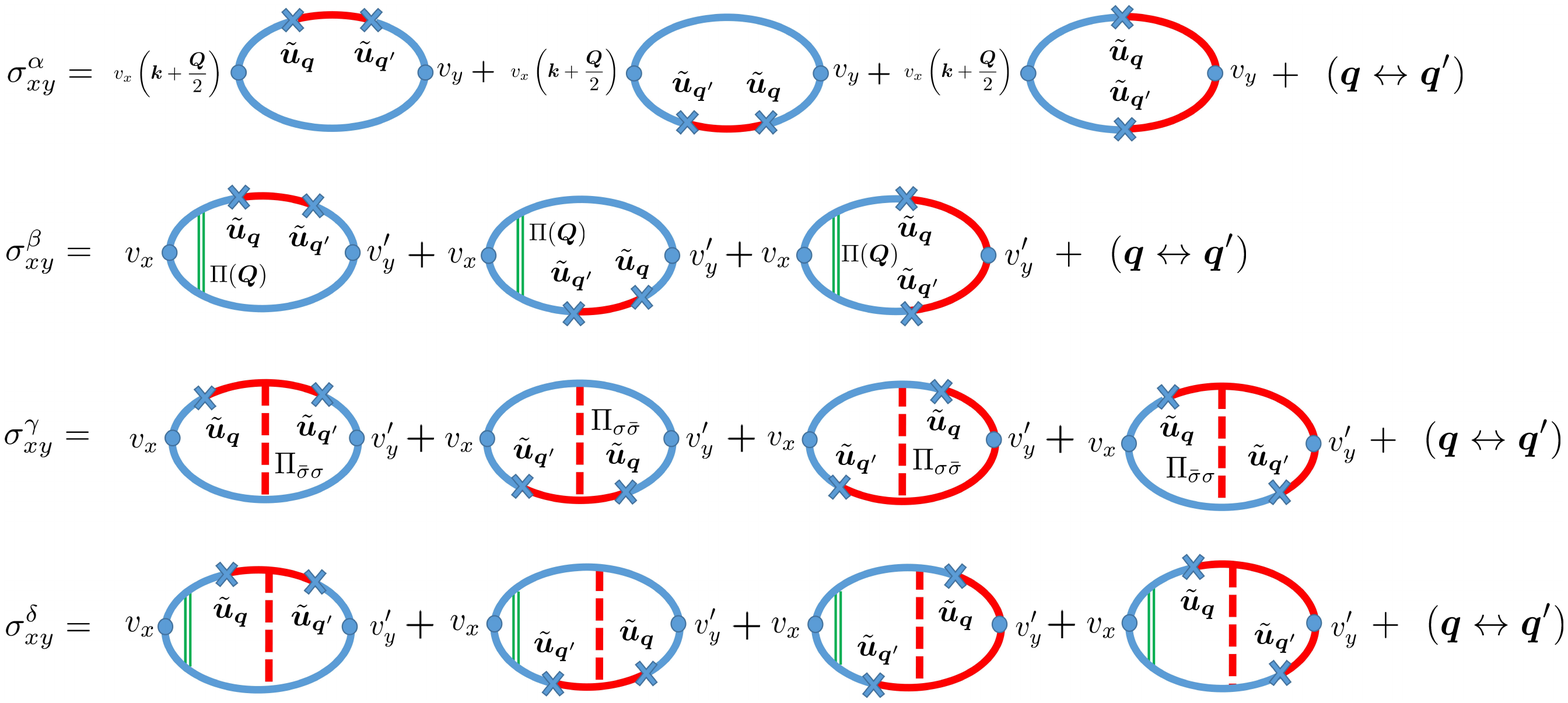}
 \end{center}
\vspace{-4.5cm}
\caption{(Color online) Feynman diagrams for THC in the $u$-perturbation calculation. 
}
 \label{fig:u_all}
\end{figure}

\subsection{Evaluation of $\sigma_{xy}^{\gamma}$ and $\sigma_{xy}^{\delta}$}

 In the following calculation, it is convenient to define 
\begin{align}
  I_{ab} &= (1+ i\zeta) \sum_{\bm k}  (G_{\bm k}^{\rm R})^{a} (G_{\bm k}^{\rm A})^{b}  , 
\\ 
  Q_{ab}^{ij} &= (1+ i\zeta) \sum_{\bm k} v_{i} v_{j} (G_{\bm k}^{\rm R})^{a} (G_{\bm k}^{\rm A})^{b}  ,  
\end{align}
where $G_{\bm k}^{\rm R \, (A)}$ is given by Eq.~(\ref{eq:green}) with $M=0$ and $\varepsilon =0$, 
and $\zeta$ by Eq.~(\ref{eq:Z-VC}).  
 We first calculate  $\sigma_{xy}^{\gamma}$, which is given by 
\begin{align}
  \sigma_{xy}^{\gamma} ({\bm Q},\omega) 
&= -\frac{e^2}{\pi} M^2 ( {\bm u}_{\bm q} \times {\bm u}_{ {\bm q}' } )^{z}  
 \sum_{\sigma} \sigma \cdot {\rm Im} \Biggl\{ X_{ {\bm q}, -{\bm q}' }^{\sigma} 
\Pi_{\bar\sigma \sigma} ( { \bm q }' ) \left[ Y_{{\bm q}'}^{\sigma}  - ( Y_{{\bm q}'}^{\bar{\sigma}} )^{*} \right]  
  - ( {\bm q} \leftrightarrow {\bm q}' ) \Biggr\} - ( x \leftrightarrow y )  , 
\end{align}
where
\begin{align}
 X_{ {\bm q}, -{\bm q}' }^{\sigma} &= (1 + i\zeta) \sum_{\bm k} \left( {\bm v} + \frac{ {\bm q} - {\bm q}' }{2m} \right)_{x}  G_{ {\bm k} + {\bm q}, \sigma }^{\rm R} G_{ {\bm k}, \bar{\sigma} }^{\rm R} G_{ {\bm k} - {\bm q}', \sigma }^{\rm A}  , 
\\
 Y_{{\bm q}}^{\sigma} &= (1 + i\zeta) \sum_{\bm k} v_{y} G_{ {\bm k} + {\bm q}, \bar{\sigma} }^{\rm R} G_{ {\bm k}, \sigma }^{\rm R} G_{ {\bm k}, \sigma }^{\rm A} .  
\end{align}
Expanding $X$ and $Y$ with respect to $q$ and $M$, this is calculated as
\begin{align}
  \sigma_{xy}^{\gamma} ({\bm Q},\omega )  
&=  - \frac{e^2 M^3}{\pi^2 \nu} \frac{( {\bm u}_{\bm q} \times {\bm u}_{ {\bm q}' } )^{z} 
    ( i{\bm q} \times i{\bm q}' )_{z}}{ \left\{ D{q'}^2\tau + (\ell/\ell_{\rm s}) \right\}^2 + (2M \tau)^2}  
									 {\rm Im} Q_{22}^{xx} \cdot {\rm Im} Q_{31}^{yy}  +  ( {\bm q} \leftrightarrow {\bm q}' )  \nonumber \\
&= - \frac{4}{9} \left( \frac{e}{m} \right)^2 \nu M^3 \tau^4 
       \frac{( {\bm u}_{\bm q} \times {\bm u}_{ {\bm q}' } )^{z} ( i{\bm q} \times i{\bm q}' )_{z}}{ \left\{ D{q'}^2\tau + (\ell/\ell_{\rm s}) \right\}^2 + (2M \tau)^2}
								+ ( {\bm q} \leftrightarrow {\bm q}' ) 
\nonumber \\
&= -\frac{4}{9} \left( \frac{e}{m} \right)^2 \nu M^3 \tau^4 \, 
   ( {\bm u}_{{\bm q}} \times {\bm u}_{ {\bm q}' } )_{z} ( i{\bm q} \times i{\bm q}')^z \,
     \left[  |\Gamma (q)|^2 + |\Gamma (q')|^2 \right]  , 
\label{eq:ugammaA} 
\end{align}
where we used 
\begin{align}
{\rm Im} Q_{22}^{xx} = {\rm Im} Q_{31}^{yy} = \frac{2 \pi}{3m} \nu \tau^2 , 
\end{align}
and ${\rm Im} I_{21}=0$. 
 We see that the dimensionless expansion parameters here are $q\ell$ and $M\tau$. 
 It should be noted that if the $Z$-VC is dropped, $Q_{22}^{xx}$ is real and 
$\sigma_{xy}^{\gamma}$ vanishes, thus inclusion of the $Z$-VC is essentially important.

 We next calculate $\sigma_{xy}^{ \delta}$, 
\begin{align}
 \sigma_{xy}^{\delta} ({\bm Q},\omega) 
&= \frac{e^2}{\pi} M^2 ( {\bm u}_{\bm q} \times {\bm u}_{ {\bm q}' } )^{z} \frac{DQ_{x}}{ (D Q^2 - i\omega ) \tau } 
 \sum_{\sigma} \sigma \cdot {\rm Re} \Biggl\{ V_{ {\bm q}, -{\bm q}' }^{\sigma} 
\Pi_{\bar\sigma \sigma} ( { \bm q }' ) \left[ Y_{{\bm q}'}^{\sigma}  - ( Y_{{\bm q}'}^{\bar{\sigma}} )^{*} \right] 
 - ( {\bm q} \leftrightarrow {\bm q}' )  \Biggr\} - ( x \leftrightarrow y )  , 
\end{align}
with
\begin{align}
 V_{ {\bm q}, -{\bm q}' }^{\sigma} (\omega) 
&= (1 + i\zeta)\sum_{\bm k} G_{ {\bm k} + {\bm q}, \sigma }^{\rm R} (\omega_{+}) G_{ {\bm k}, \bar{\sigma} }^{\rm R} (\omega_{+}) G_{ {\bm k} - {\bm q}', \sigma }^{\rm A} (\omega_{-}) ,  
\\
 Y_{{\bm q}'}^{\sigma} (\omega) 
&= (1 + i\zeta)\sum_{\bm k} v_{y} G_{ {\bm k} + {\bm q}, \bar{\sigma} }^{\rm R} (\omega_{+}) G_{ {\bm k}, \sigma }^{\rm R} (\omega_{+}) G_{ {\bm k}, \sigma }^{\rm A} (\omega_{-}) .
\end{align}
We expand $V$ and $Y$ with respect to $\omega$,
\begin{align}
  V (\omega) = V (0) - \frac{\omega}{2} V' ,  \ \ \ 
  Y (\omega) = Y (0) - \frac{\omega}{2} Y' , 
\end{align}
where $V'$ and $Y'$ are the coefficients. 
 Then we obtain
\begin{align}
 \sigma_{xy}^{\delta} ({\bm Q},\omega) 
= \  &\frac{e^2}{2\pi} M^2 ( {\bm u}_{\bm q} \times {\bm u}_{ {\bm q}' } )_{z} \frac{1}{\tau} \frac{i \omega D Q_{x}}{ DQ^2 - i\omega} 
\sum_{\sigma} \sigma \cdot {\rm Im} \Bigl\{ {V'}_{ {\bm q}, -{\bm q}' }^{\sigma} \Gamma_{ \bar{\sigma} \sigma } ( { \bm q }' ) \left[ Y_{{\bm q}'}^{\sigma}  - ( Y_{{\bm q}'}^{\bar{\sigma}} )^{*} \right] 
\nonumber \\
& \quad\quad + V_{ {\bm q}, -{\bm q}' }^{\sigma} \Gamma_{ \bar{\sigma} \sigma } ( { \bm q }' ) \left[ {Y'}_{{\bm q}'}^{\sigma}  - ( {Y'}_{{\bm q}'}^{\bar{\sigma}} )^{*} \right] 
- ( {\bm q} \leftrightarrow {\bm q}' )  
 \Bigr\}  - ( x \leftrightarrow y )   .
\end{align}
 The leading terms with respect to $M$ comes from the imaginary part of $M$-VC, giving 
\begin{align}
 \sigma_{xy}^{\delta} ({\bm Q},\omega) 
&= \frac{e^2}{2\pi} M^2 ( {\bm u}_{\bm q} \times {\bm u}_{ {\bm q}' } )_{z} \frac{1}{\tau} 
    \frac{i\omega DQ_{x}}{DQ^2 - i\omega } 
\cdot \frac{1}{2\pi \nu \tau}  \frac{8M\tau ( Dq'^2 \tau -1 )}{ (Dq'^2 + \tau_{\rm s}^{-1})^2 + (2M)^2 } \nonumber \\
&\quad \times \Bigl[ {\rm Im} V' \cdot {\rm Im} Y_{{\bm q}'} 
 + {\rm Re} V \cdot {\rm Re} Y'_{{\bm q}'} - ( {\bm q} \leftrightarrow {\bm q}' ) 
\Bigr] - ( x \leftrightarrow y) ,
\label{eq:u11}
\end{align}
where $V' \equiv ( 2 I_{31} + I_{22} )$ and $V \equiv V(0) = I_{21}$. 
 Using $I_{m1} \simeq 2\pi i \nu  (-i \tau)^{m} (1 + \zeta^2)$, 
we have ${\rm Im} V' \simeq 2\pi \nu \tau^3/(3mD)$ and ${\rm Re} V \simeq 0 \label{eq:x}$, 
which are used in Eq.~(\ref{eq:u11}). 
 After antisymmetrization, we obtain 
\begin{align}
  \sigma_{xy}^\delta ({\bm Q}, \omega)
= -\frac{4}{9} \left( \frac{e}{m} \right)^2 \nu M^3 \tau^4  
      ( {\bm u}_{{\bm q}} \times {\bm u}_{ {\bm q}' } )_{z} ( i{\bm q} \times i{\bm q}')^z 
     \left[  |\Gamma (q)|^2 + |\Gamma (q')|^2 \right]  . 
\label{eq:udeltaA} 
\end{align}
 Equations (\ref{eq:ugammaA}) and (\ref{eq:udeltaA}) lead to Eq.~(\ref{eq:ugamma}).

\subsection{On $\sigma_{xy}^{\alpha}$ and $\sigma_{xy}^{\beta}$}

 To discuss $\sigma_{xy}^{\alpha}$ and $\sigma_{xy}^{\beta}$, 
we note that the leading terms ($\sigma_{xy}^{\gamma}$ and $\sigma_{xy}^{\delta}$) depend 
on the effective spins (in regions 2 and 2$'$), which are dominant in the diffusive regime, 
or depends linearly on $M$ (in regions 1 and 1$'$). 
 Here we show that $\sigma_{xy}^{\alpha}$ and $\sigma_{xy}^{\beta}$ do not contain these features 
and thus negligible compared to $\sigma_{xy}^{\gamma}$ and $\sigma_{xy}^{\delta}$ 
in the weak-coupling diffusive region. 

 We first study $\sigma_{xy}^{\alpha}$, which has no VC. 
 A typical term is given by 
\begin{align}
& \frac{e^2}{2\pi} M^2 \sum_{\sigma} 
i\sigma ( {\bm u}_{\bm q} \times {\bm u}_{{\bm q}'} )  \sum_{\bm k} v_{i} \left( {\bm k} - \frac{\bm q}{2} - \frac{{\bm q}'}{2} \right) v_{j} G_{ {\bm k} - {\bm q} -{\bm q}', \sigma}^{\rm R} G_{ {\bm k} - {\bm q}', \bar{\sigma}}^{\rm R} G_{ {\bm k}, \sigma}^{\rm R} G_{ {\bm k}, \sigma}^{\rm A}  . 
\end{align}
 Obviously, this term does not contain the effective spin, 
since this contribution does not contain VCs. 
 Also, the expansion of the Green functions with respect to $M$ 
(i.e., $G_{\sigma} = G - \sigma M G^2 + \cdots$) is regular because of the weak-coupling condition 
($M\tau <1$).  
 This implies that the $M$-linear dependence does not appear from this term. 
 Therefore, $\sigma_{xy}^{\alpha}$ is not important in the diffusive regime.

Next, consider $\sigma_{xy}^{\beta}$, which contains one $q$-VC and is expressed as
\begin{align}
  \sigma_{ij}^{\beta} 
= \frac{e^2}{2\pi} M^2 \sum_{\sigma}  i\sigma ( {\bm u}_{\bm q} \times {\bm u}_{{\bm q}'} )^z 
   \Lambda_{1,i}^{\sigma} \Lambda_{2,j}^{\sigma}  , 
\label{eq:beta_calc}
\end{align}
with
\begin{align}
\Lambda_{1,i}^{\sigma} &= \sum_{\bm k} v_{i} G_{ {\bm k} - {\bm q}/2 - {\bm q}'/2, \sigma }^{\rm R} G_{ {\bm k} + {\bm q}/2 + {\bm q}'/2, \sigma }^{\rm A} \simeq \frac{2}{3}\pi i \nu_{\sigma} \tau_{\sigma}^2 ( q_{i} + q'_{i} ) \equiv \Lambda_{1}^{\sigma} ( q_{i} + q'_{i} ),  \\
\Lambda_{2,j}^{\sigma} &= \sum_{\bm k} v_{j} ( G^{\rm R} G^{\rm A} )_{\sigma} G_{ {\bm k} - {\bm q}', \bar{\sigma} }^{\rm R} G_{ {\bm k} - {\bm q} - {\bm q}', \sigma }^{\rm R} 
					+ \sum_{\bm k} v_{j} ( G^{\rm R} G^{\rm A} )_{\sigma} G_{ {\bm k} + {\bm q}, \bar{\sigma} }^{\rm A} G_{ {\bm k} + {\bm q} + {\bm q}', \sigma }^{\rm A} 
					+ \sum_{\bm k} v_{j} ( G^{\rm R} G^{\rm A} )_{\bar{\sigma}} G_{ {\bm k} - {\bm q}, \sigma }^{\rm R} G_{ {\bm k} + {\bm q}', \sigma }^{\rm A}.
\end{align} The term lowest order in $q$ has a factor $q^{-2}$, which may give effective spins. 
 However, since 
\begin{align}
\Lambda_{1,i}^{\sigma} \Lambda_{2,j}^{\sigma} &= \frac{2}{3} q_{i} q'_{j} \Lambda_{1}^{\sigma} {\rm Re} \left[ \sum_{\bm k} v^2  ( G_{\bar{\sigma}}^{\rm R} )^2  (G^{\rm R} G^{\rm A})_{\sigma} ( G_{\bar{\sigma}}^{\rm A} - G_{\sigma}^{\rm R}) \right]  
\end{align}
 is an even function of $M$ in the leading order with respect to $(\varepsilon_{\rm F} \tau)^{-1}$ 
 (the $\sigma$-dependent corrections are higher order), 
Eq.~(\ref{eq:beta_calc}) vanishes after the $\sigma$-sum is taken. 
 (Note that the expansion parameter is $i\sigma M\tau$, 
 hence ${\rm Re}[ \cdots ]$ is an even function of $M$.) 
 Thus the effective spins do not arise in $\sigma_{xy}^{\beta}$.
 On the other hand, terms higher order in $q$ can give finite contribution, 
but they start from $M^3 \tau^4$ and are small compared to the $M$-linear term.
 Therefore, $\sigma_{xy}^{\beta}$ is not important as well in the diffusive regime.

\section{Calculation based on $M$-perturbation}
\label{sec:M_A}

 This Appendix presents the calculation of THC based on the $M$-perturbation method. 
 Relevant diagrams are shown in Fig.~\ref{fig:Mpert}, which are  
$\sigma_{xy}^{a}$, $\sigma_{xy}^{b}$, $\sigma_{xy}^{c}$ and $\sigma_{xy}^{d}$. 
 We first study the diffusive regime (regions 1$'$ and 2$'$), and then study 
the ballistic regime (region 3). 

\begin{figure}
\hspace{0mm}
 \includegraphics[width=185mm,pagebox=cropbox,clip]{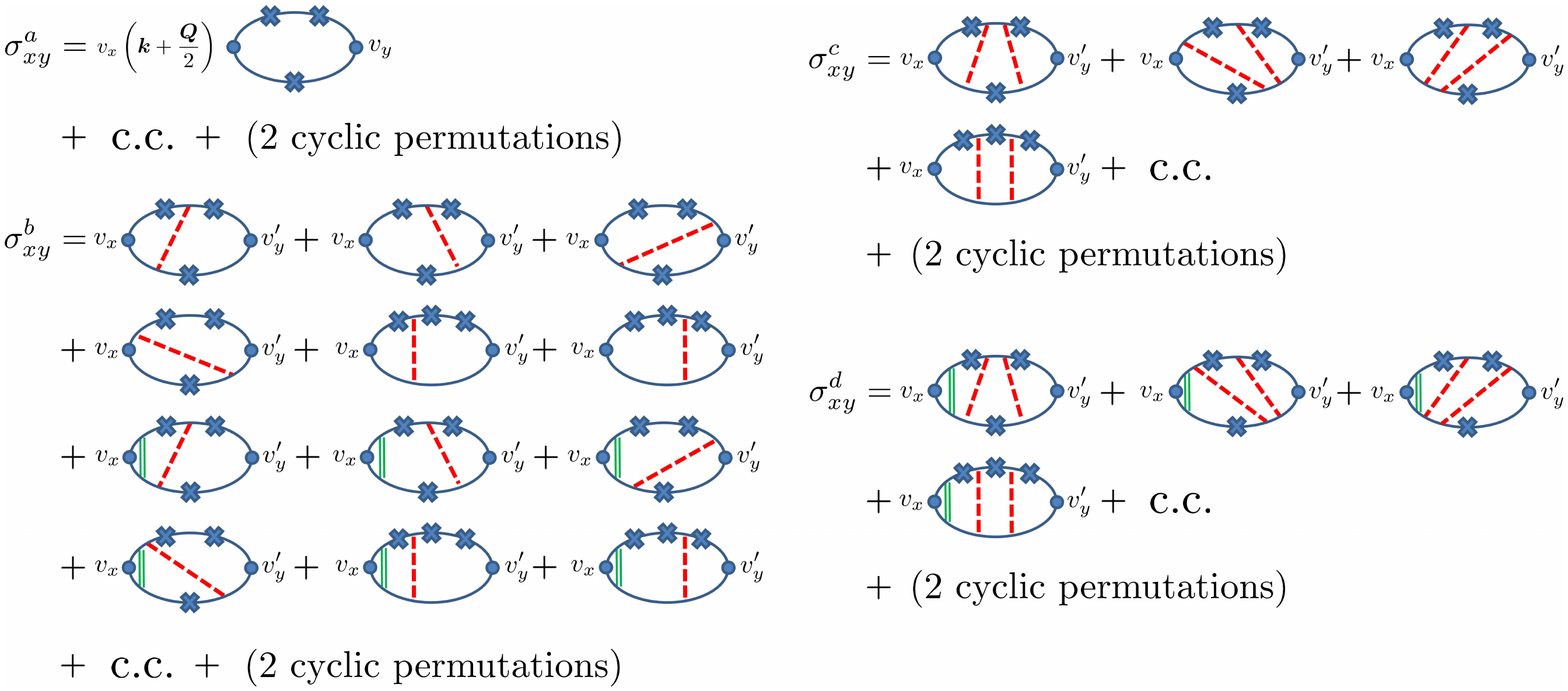}
\vspace{-4.6cm}
\caption{(Color online) Feynman diagrams for THC in the $M$-perturbation calculation. 
}
 \label{fig:M_all}
\end{figure}

\subsection{Diffusive regime (regions 1$'$ and 2$'$)}

 Calculation of $\sigma_{xy}^{a}$ without VC is straightforward, giving 
\begin{align}
  \sigma_{xy}^{a}
= -2 \left( \frac{e}{m} \right)^2 \nu M^3 \tau^4  \, 
  \langle {\bm n} \cdot \left( \partial_{x} {\bm n} \times \partial_{y} {\bm n} \right) \rangle  .  
\end{align}
 This is also obtained from Eq.~(\ref{eq:3_simple}) below. 
 
 We next consider the diagrams with the ladder type vertex corrections, 
\begin{align}
  \Pi({\bm q},\omega) = \frac{n_{\rm i}u^2}{(Dq^2 - i \omega + 1/\tau_{\rm s} ) \tau }.
\end{align}
 We have set $M=0$ since we restrict ourselves here to the purely perturbative calculation,
without taking the infinite (ladder) sum with respect to $M$.


 Focussing on the uniform component ${\bm Q}={\bm 0}$, the diagram with $q$-VC does not contribute 
to THC, similar to $\sigma_{xy}^{d}$. Diagrams with one $M$-VC are expressed as 
\begin{align}
  \sigma_{xy}^{b}
&= \frac{4e^2}{\pi} M^3  \Pi ({\bm q}) \, 
 \chi_{{\bm q},{\bm q}',{\bm q}''} \delta_{{\bm q} + {\bm q}' + {\bm q}'', {\bm 0} } \nonumber \\
  & \quad \times \Biggl[ {\rm Im} \Lambda_{1,x}({\bm q}) \ {\rm Re} \left\{  \Lambda_{2}({\bm q}', {\bm q}'') + \Lambda_{3}({\bm q}'',{\bm q}) + \Lambda_{3}^{*} ({\bm q}, {\bm q}') \right\}_{y} 
- {\rm Re} \left\{ \Lambda_{2}^{*} ({\bm q}', {\bm q}'') + \Lambda_{3}({\bm q},{\bm q}') + \Lambda_{3}^{*} ({\bm q}'', {\bm q}) \right\}_{x} \ {\rm Im} \Lambda_{1,y}({\bm q}) \Biggr] \nonumber \\
&\quad  + ( {\bm q} \leftrightarrow {\bm q}') + ({\rm 2 \ cyclic \ permutations})  , 
\label{eq:dp1}
\end{align}
with
\begin{align}
  \Lambda_{1,\alpha} ({\bm q}) 
&\equiv (1+ i\zeta) \sum_{\bm k} v_{\alpha} 
            G_{{\bm k} + {\bm q}}^{\rm R} G_{\bm k}^{\rm R} G_{\bm k}^{\rm A} 
= \frac{2\pi \nu \tau^2}{3m} \left( - 2 \epsilon_{\rm F} \tau + i \right) q_{\alpha}, \label{eq:lambda1} 
\\
\Lambda_{2,\alpha} ({\bm q}, {\bm q}')&\equiv (1+ 2i \zeta) \sum_{\bm k} v_{\alpha} G_{{\bm k} - {\bm q}}^{\rm R} G_{{\bm k} + {\bm q}'}^{\rm A} G_{\bm k}^{\rm R} G_{\bm k}^{\rm A} 
= -\frac{\pi \nu \tau^3}{m} ( {\bm q} + 3{\bm q}' )_{\alpha} + 4i \epsilon_{\rm F} \tau \frac{\pi \nu \tau^3}{m} ( {\bm q} + {\bm q}' )_{\alpha}, \label{eq:lambda2} 
\\ 
\Lambda_{3,\alpha} ({\bm q}, {\bm q}') &\equiv (1+ 2i \zeta) \sum_{\bm k} v_{\alpha} G_{{\bm k} - {\bm q}}^{\rm R} G_{{\bm k} + {\bm q}'}^{\rm R} G_{\bm k}^{\rm R} G_{\bm k}^{\rm A} 
= -\frac{\pi \nu \tau^3}{3m} \left( 1 + 4i \epsilon_{\rm F} \tau \right) ( {\bm q} - {\bm q}' )_{\alpha} . \label{eq:lambda3}
\end{align}
 Substituting Eqs.~(\ref{eq:lambda1})-(\ref{eq:lambda3}) into Eq.~(\ref{eq:dp1}), we obtain
\begin{align}
  \sigma_{xy}^{b} 
= - 48 \left( \frac{e}{m} \right)^2 \nu M^3 \tau^4 
   \left( \frac{\ell_{\rm s}}{\ell} \right)^2  \langle 
   {\bm n} \cdot \bigl( \partial_{x} \tilde {\bm n}^{(2')} \times \partial_{y} {\bm n} \bigr) \rangle 
   - (x \leftrightarrow y) , 
\label{eq_sigma_M_b}
\end{align}
where 
\begin{align}
 \tilde {\bm n}^{(2')} ({\bm r}) 
&\equiv  \frac{1}{V} \sum_{\bm q} \frac{\ell_{\rm s}^{-2}}{ q^2 + \ell_{\rm s}^{-2} } \, 
            {\bm n}_{\bm q} \, {\rm e}^{i {\bm q} \cdot {\bm r}} 
= \frac{1}{4\pi \ell_{\rm s}^2 } 
          \int d{\bm r}' \frac{ e^{- |{\bm r} - {\bm r}'|/\ell_{\rm s} } }{ |{\bm r} - {\bm r}'| } \, {\bm n} ({\bm r}') . 
\label{eq:n_tilde_2_B}
\end{align}
 Note that $\sigma_{xy}^{b}$ is proportional to $M^3 \tau^2$.
 Similarly, there are $8 \times 6$ diagrams that have two DPs, which give
\begin{align}
 \sigma_{xy}^{c} 
&= -\frac{ie^2}{\pi V} M^3 \sum_{{\bm q},{\bm q}',{\bm q}''}  \Pi ({\bm q}) \Pi ({\bm q}')  \, 
      \chi_{{\bm q},{\bm q}',{\bm q}''}
  \left[ \Lambda_{1,x} ({\bm q}) - \Lambda_{1,x}^{*} ({\bm q}) \right]  
  \left[ \Lambda_{1,y} ({\bm q}') - \Lambda_{1,y}^{*} ({\bm q}') \right] 
  \left[ \Gamma({\bm q},{\bm q}') - \Gamma^{*}({\bm q},{\bm q}') \right]
  \delta_{{\bm q} + {\bm q}' + {\bm q}'', {\bm 0} } 
\nonumber \\
&\quad  +  ( {\bm q} \leftrightarrow {\bm q}') + ({\rm 2 \ cyclic \ permutations})  , 
\end{align} 
where
\begin{align}
 \Gamma({\bm q},{\bm q}') 
&\equiv (1 + i \zeta) \sum_{\bm k} G_{{\bm k} + {\bm q}}^{\rm R} G_{{\bm k} - {\bm q}'}^{\rm R} G_{\bm k}^{\rm A}  
= -2\pi i \nu \tau^2. 
\end{align}
 This is calculated in region 1$'$ as 
\begin{align}
  \sigma_{xy}^{(1') \, c} 
= - \frac{16}{9} \left( \frac{e}{m} \right)^2 \nu M^3 \tau^2 \tau_{\rm s}^2 \, 
 \langle {\bm n} \cdot \left( \partial_{x} {\bm n} \times \partial_{y} {\bm n} \right) \rangle , 
\end{align}
 and in region 2$'$ as
\begin{align}
  \sigma_{xy}^{(2') \, c} 
= -16 \left( \frac{e}{m} \right)^2 \nu M^3 \tau^4 \, 
  \langle  {\bm n} \cdot \bigl( \partial_{x} \tilde {\bm n}^{(2')} \times \partial_{y} \tilde {\bm n}^{(2')} \bigr) \rangle , 
\label{eq:sigma_M_2d}
\end{align}
where $\tilde {\bm n}^{(2')}$ is given by Eq.~(\ref{eq:n_tilde_2_B}). 
 These lead to Eqs.~(\ref{eq:m1'}) and (\ref{eq:m2'}), respectively.

 Finally, we show that $\sigma_{xy}^{d}$ vanishes in the uniform (${\bm Q} \to {\bm 0}$) limit. 
 We first note that as ${\bm Q} \to {\bm 0}$, 
\begin{align}
 \lambda_{x} ({\bm Q}) \Pi ({\bm Q}) = \frac{1}{i\tau} \frac{DQ_{x}}{DQ^2 - i\omega} 
\to -\frac{1}{i\tau} \frac{DQ_{x}}{i\omega}  , 
\end{align}
where $\lambda_{x}$ is given by Eq.~(\ref{eq:lambda}), and 
\begin{align}
\Gamma (\omega) &= \Gamma (0) - \omega \Gamma' , 
\\
\Lambda_{y} (\omega) &= \Lambda_{y} (0) - \omega \Lambda'_{y} , 
\end{align}
where $\Gamma'$ and $\Lambda'_{y}$ are the coefficients. 
 If we retain $\omega$ in the $q$-VC, we can confirm that the $q$ dependent terms 
in $\sigma_{xy}^{d}$ vanish as ${\bm Q} \to {\bm 0}$. 
 Here we need to be careful about the terms in the numerator which are linear in $\omega$ 
since it gives a finite contribution in the limit ${\bm Q} \to {\bm 0}$.
 We focus on such \lq\lq correction'' term, $\sigma_{xy}^{d \, \prime}$, 
coming from the $\omega$ expansion of the Green functions. 
 A typical term is given by 
\begin{align}
\sigma_{xy}^{d \, \prime} ( {\bm Q} ) 
&= -i\omega \frac{8e^2}{\pi} M^3 \chi_{ {\bm q}, {\bm q}', {\bm q}'' } \lambda_{x} ({\bm Q}) \Pi ({\bm Q}) \Pi ({\bm Q}') \Pi ({\bm q}'') \nonumber \\
& \quad \times \Biggl[ 
{\rm Im} [ \Gamma' ({\bm Q}, {\bm Q}')] \cdot 
{\rm Im} [ \Gamma ( {\bm Q}', {\bm q}'' ) ]  \cdot 
{\rm Im} [ \Lambda_{y} ({\bm q}'') ]
+
{\rm Re} [ \Gamma ({\bm Q}, {\bm Q}') ] \cdot 
{\rm Re} [ \Gamma' ( {\bm Q}', {\bm q}'' ) ]  \cdot 
{\rm Im} [ \Lambda_{y} ({\bm q}'') ] 
\nonumber \\
& \quad \quad +
{\rm Re} [ \Gamma ({\bm Q}, {\bm Q}') ] \cdot 
{\rm Im} [ \Gamma ( {\bm Q}', {\bm q}'' ) ]  \cdot 
{\rm Re} [ \Lambda'_{y} ({\bm q}'') ]
\Biggr]  . 
\end{align}
 Since $\Lambda_{y}$ and $\Lambda'_{y}$ are proportional to $q''_{y}$, we see in the ${\bm Q} \to {\bm 0}$ limit that  
\begin{align}
 \sigma_{xy}^{d \, \prime} 
\propto i\omega \cdot \frac{1}{i\tau} \frac{DQ_{x}}{i\omega} \cdot 
 \chi_{ {\bm q}, {\bm q}', {\bm q}'' } ( q_{x} + q'_{x} + q''_{x} ) \, q''_{y}  , 
\end{align}
has no antisymmetric part (like ${\bm q} \times {\bm q}'$) when ${\bm Q}=0$. 
 Thus there are no \lq\lq correction'' terms in the $M$-perturbation method.

\subsection{Ballistic regime (region 3)} 

 In the ballistic regime, the dominant contribution is the one without vertex corrections. 
 They are expressed as  
\begin{align}
\sigma_{xy}^a = 
 &- \frac{i e^2}{\pi}M^3 \sum_{{\bm q},{\bm q}',{\bm q}''} 
      \chi_{{\bm q},{\bm q}',{\bm q}''} \delta_{{\bm q}+{\bm q}'+{\bm q}'',0} 
      \left( 2 I_{1,xy} - I_{2,xy} - I_{2,yx}^{*} \right) , 
\end{align}  
where $\chi_{{\bm q},{\bm q}',{\bm q}''} \equiv {\bm n}_{\bm q} \cdot \left( {\bm n}_{{\bm q}'} \times {\bm n}_{{\bm q}''} \right)$, 
\begin{align}
  I_{1,ij} &=\sum_{\bm k} v_{i} v_{j} \left( G^{\rm R} G^{\rm A} \right)_{\bm k} 
  \left( G^{\rm R} G^{\rm A} \right)_{{\bm k}-{\bm q}''} {\rm Re} \left[ G^{\rm R}_{{\bm k}+{\bm q}} \right]  , 
\\
  I_{2,ij} &=\frac{1}{m}\sum_{\bm k} v_{i} q''_{j} \left( G^{\rm R} G^{\rm A} \right)_{\bm k}  
  \left( G^{\rm R} G^{\rm A} \right)_{{\bm k}-{\bm q}''} G^{\rm R}_{{\bm k}+{\bm q}}  , 
\end{align}
with the Green function
$ G_{\bm k}^{\rm R(A)} = ( - \varepsilon_{\bm k} \pm i\gamma )^{-1}$ 
at $M=0$. 
 Here we defined $\gamma= (2\tau)^{-1}$. 
 We are interested in the antisymmetric part of the conductivity, which is contained only in $I_{2,ij}$  
but not in $I_{1,ij}$. Thus 
\begin{align}
 \sigma_{xy}^a &= 
   \frac{2e^2}{\pi} M^3 \sum_{{\bm q},{\bm q}',{\bm q}''} 
      \chi_{{\bm q},{\bm q}',{\bm q}''} \delta_{{\bm q}+{\bm q}'+{\bm q}'',0} \, 
      {\rm Im} I_{2,xy}^a , 
\end{align}  
where 
\begin{align}
  {\rm Im} I_{2,ij}^a &\equiv \frac{1}{2m}\sum_{\bm k} \left( v_{i} q''_{j} - v_{j} q''_{i} \right) 
   \left( G^{\rm R} G^{\rm A} \right)_{\bm k}  
   \left( G^{\rm R} G^{\rm A} \right)_{{\bm k}-{\bm q}''} {\rm Im} 
   \left[ G^{\rm R}_{{\bm k}+{\bm q}} \right]  . 
\end{align}
 Assuming $\gamma \ll \varepsilon_{\rm F}$ but without assuming ${\bm v} \cdot {\bm q} \gg {\bm q}^2/2m \equiv \varepsilon_{\bm q}$, 
it is calculated as 
\begin{align}
  {\rm Im} I_{2,ij}^a =& -\frac{1}{2m}\sum_{\bm k}  (v_{i} q''_{j} - v_{j} q''_{i}) \, 
 \frac{1}{(\varepsilon_{\bm k} - \varepsilon_{\rm F})^2 + \gamma^2} 
 \cdot
 \frac{1}{(\varepsilon_{\bm k} - {\bm v} \cdot {\bm q}'' + \varepsilon_{{\bm q}''}- \varepsilon_{\rm F})^2 + \gamma^2} 
 \cdot  
 \frac{\gamma}{(\varepsilon_{\bm k} + {\bm v} \cdot {\bm q} + \varepsilon_{\bm q} - \varepsilon_{\rm F})^2 + \gamma^2} 
\nonumber \\
\simeq& -\frac{\pi \nu}{m} \tau^2 \, {\rm Re} \, 
     \Biggl<\left( v_{i} q''_{j} - v_{j} q''_{i} \right) 
     \biggl[ \frac{1}{A + 2i\gamma} \frac{1}{A'' - 2i\gamma} 
             + \frac{1}{A' + 2i\gamma} \frac{1}{A - 2i\gamma} 
             + \frac{1}{A'' + 2i\gamma} \frac{1}{A' - 2i\gamma}
    \biggr] \Biggr>_{\rm FS}. 
\label{eq:I2'}
\end{align}
 The integrals in the radial direction have been evaluated as the energy integral, 
and it remains to perform the (angular) average over the Fermi surface, 
$\varepsilon_{\bm k} = \varepsilon_{\rm F}$, which is indicated by $\langle \cdots \rangle_{\rm FS}$. 
 We defined $A = {\bm v} \cdot {\bm q} + \varepsilon_{\bm q}, \ A' = {\bm v} \cdot {\bm q}' + \varepsilon_{\bm{q}''} - \varepsilon_{\bm q}, \ A'' = {\bm v} \cdot {\bm{q}''} - \varepsilon_{\bm{q}''}$. 
 Using $A + A' +A''=0$, we obtain for the Hall conductivity 
\begin{align}
  \sigma_{ij}^a ({\bm Q}={\bm 0}) 
&= \frac{2e^2}{m} \nu M^3 \tau^2 \varepsilon_{ijk} \sum_{{\bm q},{\bm q}',{\bm q}''}  
     \chi_{{\bm q},{\bm q}',{\bm q}''} \, \delta_{{\bm q}+{\bm q}'+{\bm q}'',0}  \cdot  {\rm Re} \, \Phi_{k} , 
\end{align}
or 
\begin{align}
 \sigma_{xy}^{\rm (3)} =
 \frac{2e^2}{m} \nu M^3 \tau^2 \sum_{{\bm q},{\bm q}',{\bm q}''} 
  \langle \chi_{{\bm q},{\bm q}',{\bm q}''} 
  \, {\rm e}^{i({\bm q}+{\bm q}'+{\bm q}'') \cdot {\bm r}} \, {\rm Re} \, \Phi_{z}  \rangle . 
\label{eq:formula}
\end{align}
 The information of the electron system is contained in $\Phi_{z} $, 
the $z$-component of the vector ${\bm \Phi}$ defined by  
\begin{align}
  {\bm \Phi} &= {\bm \Phi}_1 + {\bm \Phi}_2 + {\bm \Phi}_3 , 
\label{eq:Phi}
\\ 
 &{\bm \Phi}_1 =  \left\langle 
         \frac{{\bm v} \times {\bm q}}
                {({\bm v} \cdot {\bm q}'' - \varepsilon_{{\bm q}'} + \varepsilon_{\bm q} - 2i\gamma) 
                 ({\bm v} \cdot {\bm q} - \varepsilon_{\bm q} + 2i\gamma)} \right\rangle_{\rm FS} 
= {\bm F} ({\bm q}'', {\bm q} ; \varepsilon_{ {\bm q}'} - \varepsilon_{ {\bm q} }, -\varepsilon_{ {\bm q} }) \times {\bm q}  , 
\label{eq:Phi_1}
\\ 
 &{\bm \Phi}_2 =   \left\langle 
        \frac{{\bm v} \times {\bm q}'}
               {({\bm v} \cdot {\bm q}'' + \varepsilon_{{\bm q}''}- 2i\gamma)
                ({\bm v} \cdot {\bm q} + \varepsilon_{{\bm q}'} - \varepsilon_{{\bm q}''} + 2i\gamma)} \right\rangle_{\rm FS} 
= {\bm F} ({\bm q}'', {\bm q} ; -\varepsilon_{ {\bm q}'' }, \varepsilon_{ {\bm q}' } - \varepsilon_{ {\bm q}'' }) \times {\bm q}' , 
\label{eq:Phi_2}
\\
 &{\bm \Phi}_3 =  \left\langle 
        \frac{{\bm v} \times {\bm q}''}
               {({\bm v} \cdot {\bm q}'' - \varepsilon_{{\bm q}''} - 2i\gamma) 
                ({\bm v} \cdot {\bm q} + \varepsilon_{\bm q} + 2i\gamma)} \right\rangle_{\rm FS} 
= {\bm F} ({\bm q}'', {\bm q} ; \varepsilon_{ {\bm q}'' }, \varepsilon_{ {\bm q} }) \times {\bm q}'' , 
\label{eq:Phi_3}
\end{align}
where 
\begin{align}
 {\bm F} ({\bm q}, {\bm q}' ; \varepsilon, \varepsilon') 
\equiv \left< \frac{\bm v}{ ( {\bm v} \cdot {\bm q} - \varepsilon - 2i\gamma ) ( {\bm v} \cdot {\bm q}' + \varepsilon' + 2i\gamma ) } \right>_{\rm FS} , 
\label{eq:F}
\end{align}
with ${\bm v} = \hbar {\bm k}/m$ and $\varepsilon_{\bm q} \equiv \hbar^2 {\bm q}^2 /2m$. 
 The average $\langle \cdots \rangle_{\rm FS}$ can be performed by use of the Feynman's parameter integrals, with the result 
\begin{align}
 {\bm F} ({\bm q}, {\bm q}'; \varepsilon, \varepsilon') 
= \frac{ {\bm G} \times ( {\bm q} \times {\bm q}' ) }{ ( {\bm q} \times {\bm q}' )^2 } ,  
\label{eq:Fvec}
\end{align}
where
\begin{align}
 &{\bm G} = [ (\varepsilon' + 2i\gamma ){\bm q} + (\varepsilon + 2i\gamma ){\bm q}' ] 
                F ({\bm q}, {\bm q}'; \varepsilon, \varepsilon') 
                - i \left[ \, {\bm q} f (q;\varepsilon) + {\bm q}' f (q';\varepsilon') \right]  , 
\\
 &F({\bm q}, {\bm q}'; \varepsilon, \varepsilon') 
= \frac{\tau^2}{P \ell} \tan^{-1} 
    \frac{ P  \ell }{ ( 1 - i \varepsilon \tau ) ( 1 - i \varepsilon' \tau ) - \ell^2 {\bm q} \cdot {\bm q}' }  , 
\label{eq:F-f}
 \\
 &f(q, \varepsilon) 
\equiv \left< \frac{i}{ {\bm v} \cdot {\bm q} + \varepsilon + 2i\gamma } \right>_{\rm FS} 
= \frac{\tau}{q\ell} \tan^{-1} \frac{q\ell}{1- i\varepsilon \tau}  , 
\label{eq:f-f}
\end{align}
with $P  = \sqrt{ [ (1 - i \varepsilon' \tau){\bm q} + (1 - i \varepsilon \tau){\bm q}' ]^2 + \ell^2 ( {\bm q} \times {\bm q}' )^2 }$. 
 The above expression is valid as far as $\gamma \ll \varepsilon_{\rm F}$ and $q \ll k_{\rm F}$, 
irrespective of the relative magnitude of $\varepsilon_{\bm q}$ and $\gamma$.

\subsection{More on simple bubble diagram}

 The results for simple bubble diagrams (i.e., without vertex corrections) 
obtained in the preceding subsection are quite general 
(only assumptions made are $\gamma \ll \varepsilon_{\rm F}$ and $q \ll k_{\rm F}$), 
but show complicated appearance. 
 However, they can be simplified if  $\varepsilon_q , \varepsilon_{q'}, \varepsilon_{q''} \ll \gamma$, 
where $\varepsilon_{q}  = \hbar^2 q^2/2m$. 
 In this case, one can make a shift ${\bm k} \to {\bm k} + {\bm q}$ of the momentum ${\bm k}$ 
to evaluate ${\bm F}$ in Eq.~(\ref{eq:F}). 
 This is because the Fermi surface is smeared by the damping $\gamma$ and the 
change of energy associated with the above shift is negligible compared to $\gamma$.

 To be explicit, we write Eqs.~(\ref{eq:Phi_1})-(\ref{eq:Phi_3}) as 
\begin{eqnarray}
 {\bm \Phi}_1 &=&  \left\langle 
         \frac{{\bm v} \times {\bm q}}
                {( {\bm q}'' \cdot ({\bm v} + \frac{{\bm q}'-{\bm q}}{2m} ) - 2i\gamma) 
                 ( {\bm q} \cdot ({\bm v} - \frac{{\bm q}}{2m} ) + 2i\gamma)} \right\rangle_{\rm FS}  ,      
\\
 {\bm \Phi}_2 &=&  \left\langle 
        \frac{{\bm v} \times {\bm q}'}
               {( {\bm q}'' \cdot ({\bm v} + \frac{{\bm q}''}{2m} ) - 2i\gamma)
                ( {\bm q} \cdot ({\bm v} + \frac{{\bm q}''-{\bm q}'}{2m} ) + 2i\gamma)} \right\rangle_{\rm FS}  \\
 {\bm \Phi}_3 &=& \left\langle 
        \frac{{\bm v} \times {\bm q}''}
               {( {\bm q}'' \cdot ({\bm v} - \frac{{\bm q}''}{2m} ) - 2i\gamma) 
                ( {\bm q} \cdot ({\bm v} + \frac{{\bm q}}{2m} ) + 2i\gamma)} \right\rangle_{\rm FS}  , 
\end{eqnarray}
and make a shift as ${\bm v} \to {\bm v} + \frac{{\bm q}}{m}$ in ${\bm \Phi}_1$, 
and ${\bm v} \to {\bm v} - \frac{{\bm q}''}{m}$ in ${\bm \Phi}_2$, 
 Then 
\begin{eqnarray}
 {\bm \Phi}_1 &\to&   \left\langle 
         \frac{{\bm v} \times {\bm q}}
                {( {\bm q}'' \cdot ({\bm v} + \frac{{\bm q}+{\bm q}'}{2m} ) - 2i\gamma) 
                 ( {\bm q} \cdot ({\bm v} + \frac{{\bm q}}{2m} ) + 2i\gamma)} \right\rangle_{\rm FS} 
\ = \  \left\langle 
         \frac{{\bm v} \times {\bm q}}
                {( {\bm q}'' \cdot ({\bm v} - \frac{{\bm q}''}{2m} ) - 2i\gamma) 
                 ( {\bm q} \cdot ({\bm v} + \frac{{\bm q}}{2m} ) + 2i\gamma)} \right\rangle_{\rm FS} , 
\\
 {\bm \Phi}_2 &\to& \left\langle 
        \frac{({\bm v} - \frac{{\bm q}''}{m}) \times {\bm q}'}
               {( {\bm q}'' \cdot ({\bm v} - \frac{{\bm q}''}{2m} ) - 2i\gamma)
                ( {\bm q} \cdot ({\bm v} - \frac{{\bm q}'+{\bm q}''}{2m} ) + 2i\gamma)} \right\rangle_{\rm FS}   
\ = \  \left\langle 
        \frac{({\bm v} - \frac{{\bm q}''}{m}) \times {\bm q}'}
               {( {\bm q}'' \cdot ({\bm v} - \frac{{\bm q}''}{2m} ) - 2i\gamma)
                ( {\bm q} \cdot ({\bm v} + \frac{{\bm q}}{2m} ) + 2i\gamma)} \right\rangle_{\rm FS}  , 
\end{eqnarray}
where we used ${\bm q} + {\bm q}' + {\bm q}'' = {\bm 0}$. 
 Therefore, 
\begin{eqnarray}
  {\bm \Phi}  \ = \  {\bm \Phi}_1 + {\bm \Phi}_2 + {\bm \Phi}_3
&=&  \frac{{\bm q}' \times {\bm q}'' }{m}  \left\langle 
        \frac{1}{( {\bm q}'' \cdot ({\bm v} - \frac{{\bm q}''}{2m} ) - 2i\gamma)
                   ( {\bm q} \cdot ({\bm v} + \frac{{\bm q}}{2m} ) + 2i\gamma)} \right\rangle_{\rm FS} . 
\end{eqnarray}
 Because of the overall factor of ${\bm q}' \times {\bm q}''$, 
we may neglect $\varepsilon_{\bm q}$ and $\varepsilon_{\bm q''}$ in the denominator to obtain 
\begin{eqnarray}
 {\bm \Phi} 
&\simeq&   \frac{{\bm q} \times {\bm q}' }{m}  \left\langle 
        \frac{1}{( {\bm q}'' \cdot {\bm v} - 2i\gamma) ( {\bm q} \cdot {\bm v} + 2i\gamma)} \right\rangle_{\rm FS}  
\ = \   \frac{{\bm q} \times {\bm q}' }{m}  F ( {\bm q}'', {\bm q} ; \, 0, \, 0  ) . 
\end{eqnarray}
  At $\varepsilon = \varepsilon' = 0$, the function $F$ has a relatively simple form, 
\begin{eqnarray}
 F ({\bm q}, {\bm q}'; 0, 0) 
&=&  \frac{\tau^2}{P \ell} \, {\rm tan}^{-1} \left[ \frac{P \ell} {1 - \ell^2 \, {\bm q} \cdot {\bm q}'} \right] , 
\ \ \ \ \ \ \  
 P \ = \ \sqrt{ ({\bm q}+{\bm q}')^2 + \ell^2 ({\bm q} \times {\bm q}')^2 } , 
\end{eqnarray}
and we obtain 
\begin{eqnarray}
 {\bm \Phi} 
&\simeq&  \frac{{\bm q} \times {\bm q}' }{m}  
   \frac{\tau^2}{P \ell} \, {\rm tan}^{-1} \left[ \frac{P \ell} {1 - \ell^2 \, {\bm q} \cdot {\bm q}''} \right]  
\ = \   \frac{{\bm q} \times {\bm q}' }{m}  
   \frac{\tau^2}{P \ell} \, {\rm tan}^{-1} \left[ \frac{P \ell} {1 + \ell^2 q (q + q' \cos \theta_{{\bm q}, {\bm q}'}) } \right]  , 
\label{eq:Phi_simple}
\end{eqnarray}
with
\begin{eqnarray}
 P &=&\sqrt{ q^{\prime \, 2} + \ell^2 ({\bm q} \times {\bm q}')^2 } 
   \ = \  q' \sqrt{ 1 + ( q \ell \sin \theta_{{\bm q}, {\bm q}'})^2 } ,
\end{eqnarray}
where $\theta_{{\bm q}, {\bm q}'}$ is the angle between ${\bm q}$ and ${\bm q}'$. 

 Therefore, when $\varepsilon_q \ll \gamma$, we obtain a rather simple expression, 
\begin{align}
  \sigma_{xy}^a ({\bm Q}={\bm 0}) 
&= - 2 \left( \frac{e}{m} \right)^2 \nu M^3 \tau^4 \sum_{{\bm q},{\bm q}',{\bm q}''} 
     (i{\bm q} \times i{\bm q}')^z  \chi_{{\bm q},{\bm q}',{\bm q}''}  
     \frac{1}{P \ell} \, {\rm tan}^{-1} \left[ \frac{P \ell} {1 - \ell^2 \, {\bm q} \cdot {\bm q}''} \right] 
      \delta_{{\bm q}+{\bm q}'+{\bm q}'',0} , 
\label{eq:3_simple}
\end{align}
for a simple bubble diagram of THC.

\section{Calculation for skyrmion lattice}

 In this Appendix, we apply the results obtained by $M$-perturbation for general spin textures 
to a skyrmion lattice. 
 We consider a triple-$q$ state, and focus on the \lq\lq weakest-coupling'' regions, 1$'$, 2$'$ and 3. 
 Under the assumed conditions, $\gamma \ll \varepsilon_{\rm F}$ and $q \ll k_{\rm F}$, 
it is convenient to consider the following three regions, 
\begin{align}
	{\rm I.} &\ \  q\ell < 1 < \sqrt{k_{\rm F} \ell} < k_{\rm F}\ell  \ \  
	  (\varepsilon_{q} < v_{\rm F}q < \gamma < \varepsilon_{\rm F})  ,  
\nonumber \\
	{\rm II.} &\ \  1 < q\ell < \sqrt{k_{\rm F} \ell} < k_{\rm F}\ell  \ \  
	  (\varepsilon_{q} < \gamma < v_{\rm F}q < \varepsilon_{\rm F})  , 
\nonumber \\
	{\rm III.} &\ \  1 < \sqrt{k_{\rm F} \ell} < q\ell < k_{\rm F}\ell  \ \  
    (\gamma < \varepsilon_{q} < v_{\rm F}q < \varepsilon_{\rm F}) .
\end{align}
 The diffusive region I contains regions 1$'$ and 2$'$ in the text, and the ballistic regions II and III 
correspond to region 3. 
 For regions I and II, we can use the simple expression (\ref{eq:3_simple}) for THC, 
and we can discuss the crossover between the diffusive and ballistic regimes. 
 Note, however, that the THC in region I is dominated by diagrams with vertex corrections.

 The THC for the triple-$q$ state is expressed as 
\begin{align}
  \sigma_{xy} \bigr|_{\rm skL} =
\frac{4e^2}{m} \nu M^3 \tau^2 \sum_{\ell, m, n} {\rm Re}[ \chi_{\ell,m,n} ]  \, 
  \delta_{{\bm q}_\ell +{\bm q}_m +{\bm q}_n ,0} \, {\rm Re} \, \Phi_z  , 
\end{align}
where $\ell, m,n$ take $a,b$ and $c$. 
 Note that a factor of two comes from the complex conjugate terms in Eq.~(\ref{eq:M_sk}). 
 ${\bm \Phi}$ is given by Eq.~(\ref{eq:Phi}) or (\ref{eq:Phi_simple}), 
with $\varepsilon_{q_i} = \varepsilon_{q_j}$, 
${\bm q}_i \cdot {\bm q}_i = -q^2/2$, 
and $ | {\bm q}_i \times {\bm q}_j | = \sqrt{3} \, q^2/2$ for $i \ne j$.

\subsection{Region I ($q\ell < 1 < \sqrt{k_{\rm F} \ell} < k_{\rm F}\ell$ 
 or $\varepsilon_{q} < v_{\rm F}q < \gamma < \varepsilon_{\rm F})$}

In this region, we can use Eq.~(\ref{eq:3_simple}), 
\begin{align}
  \sigma_{xy}^a \Bigr|_{\rm skL}  
= -4\left( \frac{e}{m} \right)^2 \nu M^3 \tau^4  
    \frac{1}{ q\ell \sqrt{ 1 + \frac{3}{4} (q\ell)^2 } } 
    \tan^{-1} \left[ \frac{ q \ell \sqrt{ 1 + \frac{3}{4} (q\ell)^2 } }{ 1 + \frac{1}{2} (q\ell)^2 } \right] 
    \sum_{a,b,c} {\rm Re}[ \chi_{a,b,c} ] (i {\bm q}_a \times i {\bm q}_b )_z  . 
\label{eq:THC_skL_I,II}
\end{align}
 Considering the case of small $q\ell$, we obtain 
\begin{align}
  \sigma_{xy}^{{\rm (I)} , \, a} \Bigr|_{\rm skL} 
= -4\left( \frac{e}{m} \right)^2 \nu M^3 \tau^4  
      \sum_{a,b,c} {\rm Re}[ \chi_{a,b,c} ] (i {\bm q}_a \times i {\bm q}_b )_z  . 
\end{align}
 However, since this is in the diffusive regime, $q\ell <1$, 
the contribution with ladder vertex correction will be dominant. 
 Similarly to Eqs.~(\ref{eq_sigma_M_b}), (\ref{eq:n_tilde_2_B}) and (\ref{eq:sigma_M_2d}), we obtain 
\begin{align}
  \sigma_{xy}^{{\rm (I)}, \, c} \Bigr|_{\rm skL} 
=  - 16 \left( \frac{e}{m} \right)^2 \nu M^3 \tau^4 
   \frac{1}{[(q\ell)^2 + (\ell/\ell_{\rm s})^2]^2}  \sum_{a, b, c} {\rm Re}[ \chi_{a,b,c} ]
   \left( i{\bm q}_a \times i{\bm q}_b  \right)_z , 
\label{eq:THC_skL_I} 
\end{align}
and this is indeed the main contribution in region I.

\subsection{Region II ($1 < q\ell < \sqrt{k_{\rm F} \ell} < k_{\rm F}\ell$ 
 or $\varepsilon_{q} < \gamma < v_{\rm F}q < \varepsilon_{\rm F})$}

 In this region, we can still use Eq.~(\ref{eq:THC_skL_I,II}), but consider the case, $q\ell > 1$. 
Then, 
\begin{align}
 \sigma_{xy}^{\rm (II)} \Bigr|_{\rm skL} 
&= -\frac{8\sqrt{3}\pi}{9} \left( \frac{e}{m} \right)^2 \nu M^3 \tau^4 \frac{1}{(q \ell)^2}
 \sum_{a,b,c} {\rm Re}[ \chi_{a,b,c} ] ( i {\bm q}_a \times i {\bm q}_b )_z . 
\label{eq:THC_skL_II} 
\end{align}

\subsection{Region III ($1 < \sqrt{k_{\rm F} \ell} < q\ell < k_{\rm F}\ell$ 
 or $\gamma < \varepsilon_{q} < v_{\rm F}q < \varepsilon_{\rm F}$)}

 In this region, we need to use the general expression, Eqs.~(\ref{eq:formula})-(\ref{eq:f-f}). 
 From these, ${\bm \Phi}_n$'s are given by 
\begin{align}
{\bm \Phi}_1 &= {\bm F} ({\bm q}_c,{\bm q}_a;0,-\varepsilon_{q}) \times {\bm q}_a 
\equiv {\bm F}_a \times {\bm q}_a , 
\\
{\bm \Phi}_2 &= {\bm F} ({\bm q}_c,{\bm q}_a;-\varepsilon_{q},0) \times {\bm q}_b 
\equiv {\bm F}_b \times {\bm q}_b , 
\\
{\bm \Phi}_3 &= {\bm F} ({\bm q}_c,{\bm q}_a;\varepsilon_{q}, \varepsilon_{q}) \times {\bm q}_c 
\equiv {\bm F}_c \times {\bm q}_c , 
\end{align}
where $|{\bm q}_a| =  |{\bm q}_b| =  |{\bm q}_c| \equiv q$, and we defined ${\bm F}_\ell$. 
 Also defining ${\bm G}_\ell$ by 
${\bm F}_\ell = {\bm G}_\ell \times ({\bm q}_c \times {\bm q}_a)/|{\bm q}_c \times {\bm q}_a|^2 $, 
see Eq.~(\ref{eq:Fvec}), 
one can show that 
\begin{align}
  {\bm \Phi} &= \frac{{\bm q}_a \times {\bm q}_b}{|{\bm q}_a \times {\bm q}_b|^2}  
                      \sum_{\ell = a,b,c} ({\bm G}_\ell\cdot {\bm q}_\ell )  . 
\end{align}
 The ${\bm G}_\ell$'s are given by 
\begin{align}
  {\bm G}_a  &= \left[ \, (-\varepsilon_q + 2i\gamma) {\bm q}_c + 2i \gamma {\bm q}_a  \right] F_a 
  - i \left[ \, {\bm q}_c f(q;0) + {\bm q}_a f(q; -\varepsilon_q)  \right]  , 
\\
  {\bm G}_b  &= \left[ \,  2i \gamma {\bm q}_c +  (-\varepsilon_q + 2i\gamma) {\bm q}_a \right] F_b 
  - i \left[ \, {\bm q}_c f(q; -\varepsilon_q) + {\bm q}_a f(q; 0)  \right]  , 
\\
  {\bm G}_c  &= ({\bm q}_a + {\bm q}_c) 
    \left[ \, (\varepsilon_q + 2i\gamma)F_c - i f(q; \varepsilon_q)  \right] , 
\end{align}
where $F_\ell$'s are $F$ functions [Eq.~(\ref{eq:F-f})] having the same arguments 
as ${\bm F}_\ell$, namely, $F_a = F({\bm q}_c,{\bm q}_a;0,-\varepsilon_{q})$, etc. 
 Using ${\bm q}_i \cdot {\bm q}_j = -q^2/2$ for $i \ne j$, one has 
\begin{align}
  {\bm G}_a \cdot {\bm q}_a 
&= \frac{1}{2} q^2  \left\{ (\varepsilon_q + 2i\gamma) F_a 
  - i [ \, 2 f(q; -\varepsilon_q) - f(q;0) ] \right\} , 
\\
  {\bm G}_b \cdot {\bm q}_b
&=  \frac{1}{2} q^2 \left\{  
    (\varepsilon_q - 4i \gamma ) F_b + i [ \, f(q; -\varepsilon_q) + f(q; 0) ] \right\}  , 
\\
  {\bm G}_c \cdot {\bm q}_c
&= \frac{1}{2} q^2 \left\{ (\varepsilon_q + 2i\gamma)F_c - i f(q; \varepsilon_q)  \right\}  , 
\end{align}
and thus 
\begin{align}
  {\rm Re} \sum_{\ell = a,b,c} ({\bm G}_\ell\cdot {\bm q}_\ell )  
&=  \frac{1}{2} q^2 \,  {\rm Re} 
      \left\{  \varepsilon_q (F_a + F_b + F_c ) + 2i \gamma (F_a - 2F_b + F_c ) \right\}  .  
\end{align}
 By retaining the leading term under the present condition, $\sqrt{k_{\rm F} \ell} < q\ell $ 
or $\gamma < \varepsilon_{q}$, one may estimate as 
$P\ell  \simeq \ell^2 \, |{\bm q}_a \times {\bm q}_b| =  \sqrt{3} \, (q \ell )^2/2$, 
thus 
\begin{align}
  F_a  \simeq  F_b  \simeq  F_c  
\simeq \frac{\tau^2}{P \ell} \, {\rm tan}^{-1} \sqrt{3} 
\simeq   \frac{2\pi}{3\sqrt{3}}\frac{1}{(qv_{\rm F})^2}  . 
\end{align}
 This leads to  
\begin{align}
 {\rm Re} \, {\bm \Phi} 
&= \frac{ 2 \sqrt{3} \pi}{9m} \cdot \frac{{\bm q}_a \times {\bm q}_b}{(q v_{\rm F})^2}   
= \frac{ 2 \sqrt{3} \pi}{9m} \left( \frac{\tau}{q\ell} \right)^2 ({\bm q}_a \times {\bm q}_b) , 
\end{align}
and the THC is obtained as 
\begin{align}
  \sigma_{xy}^{\rm (III)} \Bigr|_{\rm skL} 
&= - \frac{ 8 \sqrt{3} \pi}{9}  \left( \frac{e}{m} \right)^2 
   \nu M^3 \tau^4 \frac{1}{(q \ell)^2}  
   \sum_{\ell, m, n} {\rm Re}[ \chi_{\ell,m,n} ]  \, 
   (i{\bm q}_a \times i{\bm q}_b) \,   \delta_{{\bm q}_\ell +{\bm q}_m + {\bm q}_n ,0} 
\label{eq:THC_skL_III} 
\\ 
&=\sigma_{xy}^{\rm (II)} \Bigr|_{\rm skL}  .  
\end{align}
 This result indicates that the scale $\sqrt{k_{\rm F}\ell}$ 
(or the energy scale difference between $\varepsilon_q$ and $\gamma$) 
is not important for the THE in a skyrmion lattice. 
 However, we note that this scale difference may be important for general spin textures  
since the $F$ function is sensitive to the angle between ${\bm q}$ and ${\bm q}'$, etc.

\section{Relation between Denisov {\it et al.} and TK}

 In this Appendix, we study the relation between Denisov {\it et al.} \cite{Denisov} and TK \cite{Tatara}. 
 We consider large momentum transfer processes of electrons at the scattering by spin texture 
(like magnetic impurities), and attempt to reproduce the result (\ref{eq:denisov}) from our standpoint. 
 We consider the $s$-$d$ coupling of the form, 
\begin{align}
 H_{sd}^{{\rm large} -q} 
= - \frac{M}{V} \sum_{{\bm k},{\bm k}'} 
   \sum_{j} {\bm m}_{j} \cdot \left( c_{\bm k}^{\dagger} {\bm \sigma} c_{{\bm k}'} \right) 
   {\rm e}^{i ({\bm k}' - {\bm k}) \cdot {\bm R}_{j}} , 
\end{align}
where, ${\bm R}_{j}$ is the position of the localized moment ${\bm m}_{j}$, 
and calculate the Hall conductivity by the third order perturbation, 
\begin{align}
\sigma_{xy}^{\rm T'} &= -\frac{e^2}{2\pi} M^3 \int d{\bm r} d{\bm r}' d{\bm R}_{1} d{\bm R}_{2} d{\bm R}_{3} \  {\rm Im} G^{\rm R} ({\bm R}_{1}, {\bm R}_{2}) \ [ {\bm m}_{1} \cdot ( {\bm m}_{2} \times {\bm m}_{3} ) ]
\nonumber \\
&\quad \times \Biggl[ (\partial_{x} - \partial_{x}') G^{\rm R} ({\bm R}_{3}, {\bm r}) G^{\rm A} ({\bm r}', {\bm R}_{1}) \cdot (\partial_{y} - \partial_{y}') G^{\rm R} ({\bm R}_{2}, {\bm r}) G^{\rm A} ({\bm r}', {\bm R}_{3}) \Biggr]  .
\end{align}
  Going to the Fourier representation and evaluating as 
$\sum_{\bm k} G^{\rm R}_{\bm k} G^{\rm A}_{\bm k} \sim \nu \tau$, 
as coming from states near the Fermi surface 
(leaving the angular dependence of the phase factor to $\Omega$), we obtain
\begin{align}
\sigma_{xy}^{\rm T'} \simeq \sigma_{0} \Omega \, \tau \simeq \sigma_{xy}^{\rm T}  , 
\end{align}
where $\sigma_{0} = e^2 n\tau/m$ is the Boltzmann conductivity, $\Omega$ is proportional to 
\begin{align}
\Omega \propto &-M^3 \ {\rm Im} \int d{\bm r}_{1} d{\bm r}_{2} d\theta \sin{\theta} \ G^{\rm R} ({\bm r}_{1}, {\bm r}_{2}) \  {\rm e}^{-i({\bm p}_{2} \cdot {\bm r}_{1} + {\bm p}_{1} \cdot {\bm r}_{2})} 
 \int d{\bm r} \ {\bm m} ({\bm r}) \cdot ( {\bm m} ({\bm r} + {\bm r}_{1}) \times {\bm m} ({\bm r} + {\bm r}_{2}) )  , 
\end{align}
with 
${\bm r} = {\bm R}_{3}$, ${\bm r}_{1} = {\bm R}_{1} - {\bm R}_{3}$, ${\bm r}_{2} = {\bm R}_{2} - {\bm R}_{3}$, 
and $\theta$ is the angle between ${\bm p}_{1}$ and ${\bm p}_{2}$. 
 This reproduces the leading term of Eq.~(\ref{eq:denisov}), meaning that the situation 
considered by Denisov {\it et al.} \cite{Denisov} corresponds to discretely distributed spin system    
as studied by TK.

\end{widetext}


\begin{thebibliography}{9}
\bibitem{com1} Strictly speaking, $M$-perturbation is applicable in the case 
$M\tau_{\rm s} < (q \ell_{\rm s})^2 < 3$, however, region 1$'$ is characterized by the relation 
$ (q \ell_{\rm s})^2 + M\tau_{\rm s} < 3$, hence it is included in the criterion of region 1$'$.
\end{thebibliography}
\end{document}